\documentclass[letterpaper,conference,onecolumn,11pt]{IEEEtran}
\IEEEoverridecommandlockouts                              
\usepackage{srcltx}
\usepackage{color}
\usepackage{graphicx}
\usepackage{bm}
\usepackage[cmex10]{amsmath}
\usepackage{algorithmic}
\usepackage{algorithm}
\usepackage{amssymb}
\usepackage{subfigure}
\usepackage{multirow}
\newtheorem{lemma}{Lemma}
\newtheorem{remark}{Remark}

\newtheorem{assumption}{Assumption}

 \newtheorem{pf}{Proof}
\usepackage{amsfonts}

\begin{document}

\title{Learning-based Reduced Order Model Stabilization for Partial Differential Equations: Application to the Coupled Burgers Equation}
\author{Mouhacine Benosman, Boris Kramer, Petros Boufounos, Piyush Grover
\thanks{ M. Benosman
(m{\_}benosman@ieee.org) is with the Multimedia group of
Mitsubishi Electric Research Laboratories (MERL), Cambridge, MA
02139, USA. B. Kramer is with the Department of Aeronautics and
Astronautics,  Massachusetts Institute of Technology, Cambridge,
MA, 02139. P. Boufounos is with the Multimedia group of MERL. P.
Grover is with the Mechatronics group of MERL}}

\maketitle

\begin{abstract}
We present results on stabilization for reduced order models (ROM)
of partial differential equations using learning. Stabilization is
achieved via closure models for ROMs, where we use a model-free
extremum seeking (ES) dither-based algorithm to learn the best
closure models' parameters, for optimal ROM stabilization. We
first propose to auto-tune linear closure models using ES, and
then extend the results to a closure model combining linear and
nonlinear terms, for better stabilization performance. The coupled
Burgers' equation is employed as a test-bed for the proposed
tuning method.
\end{abstract}

\section{Introduction}
\label{intro} The problem of reducing a partial differential
equation (PDE) model to a system of finite dimensional ordinary
differential equations (ODE), has significant applications in
engineering and physics, where solving such PDE models is too time
consuming. Reducing the PDE model to a simpler representation,
without loosing the main characteristics of the original model,
such as stability and prediction precision, is appealing for any
real-time model-based computations. However, this problem remains
challenging, since model reduction can introduce stability loss
and prediction degradation. To remedy these problems, many methods
have been developed aiming at what is known as stable model
reduction.

In this paper, we focus on additive terms called {\em closure
models} and their application in reduced order model (ROM)
stabilization. We develop a learning-based method, applying
extremum-seeking (ES) methods to automatically tune the
coefficients of the closure models and obtain an optimal
stabilization of the ROM.

Our work extends some of the existing results in the field. For
instance, a reduced order modelling method is proposed in
\cite{CNTLDBDN13} for stable model reduction of Navier-Stokes flow
models. The authors propose stabilization by adding a nonlinear
viscosity stabilizing term to the reduced order model. The
coefficients of this term are identified using a variational
data-assimilation approach, based on solving a deterministic
optimization. In \cite{BDN13,B13}, a Lyapunov-based stable model
reduction is proposed for incompressible flows. The approach is
based on an iterative search of the projection modes satisfying a
local Lyapunov stability condition.

An example of stable model reduction for the Burger's equation
using closure models is explored in~\cite{SB14,SI13}. These
closure models modify some stability-enhancing coefficients of the
reduced order ODE model using either constant additive terms, such
as the constant eddy viscosity model, or time and space varying
terms, such as Smagorinsky models. The added terms' amplitudes are
tuned in such a way to stabilize the reduced order model. However,
such tuning is not always straightforward. Our work addresses this
issue and achieves optimal tuning using learning-based approaches.

This paper is organized as follows: Section \ref{prem} establishes our
notation and some necessary definitions. Section \ref{es-rom-stab}
introduces the problem of PDE model reduction and the closure
model-based stabilization, and presents the main result of this paper.
An example using the coupled Burgers' equation is treated in Section
\ref{Burgers-example}.  Finally, Section \ref{concl} provides some
discussion on our approach and concludes.

\section{Basic Notations and definitions}
\label{prem}
Throughout the paper we will use $\|.\|$ to denote the Euclidean
vector norm; i.e., for $x\in \mathbb{R}^n$ we have $\|x\|=\sqrt{x^T
  x}$. The Kronecker delta function is defined as:
$\delta_{ij}=0,\;\text{for}\;i\neq j$ and $\delta_{ii}=1$. We will
use $\dot{f}$ for the short notation of time derivative of $f$,
and $x^{T}$ for the transpose of a vector $x$. A function is said
analytic in a given set, if it admits a convergent Taylor series
approximation in some neighborhood of every point of the set. We
consider the Hilbert space $\mathcal{Z}=L^{2}([0,1])$, which is
the space of Lebesgue square integrable functions, i.e.,
$f\in\mathcal{Z}$, iff $\int_{0}^{1}|f(x)|^{2}dx<\infty$. We
define on $\mathcal{Z}$ the inner product $\langle
\cdot,\cdot\rangle _{\mathcal{Z}}$ and the associated norm
$\|.\|_{\mathcal{Z}}$, as
$\|f\|_{\mathcal{Z}}^{2}=\int_{0}^{1}|f(x)|^{2}dx$, and $\langle
f,g\rangle _{\mathcal{Z}}=\int_{0}^{1}f(x)g(x)dx$, for
$f,g\in\mathcal{Z}$. A function $\omega(t,x)$ is in
$L^{2}([0,T];\mathcal{Z})$ if for each $0\leq t\leq T$,
$\omega(t,\cdot)\in\mathcal{Z}$, and
$\int_{0}^{T}{\|\omega(t,\cdot)\|_\mathcal{Z}^2} dt\leq\infty$.
Finally, in the remaining of this paper by stability we mean
stability of dynamical systems in the sense of Lagrange, e.g.,
\cite{haddad2008}.

\section{ES-based PDEs stable model reduction}
\label{es-rom-stab} We consider a stable dynamical system modelled
by a nonlinear partial differential equation of the form
\begin{equation}\label{general_PDE_chap3}
\dot{z} = \mathcal{F}(z) \in\mathcal{Z},
\end{equation}
where $\mathcal{Z}$ is an infinite-dimension Hilbert space.
Solutions to this PDE can be obtained through numerical
discretization, using, e.g., finite elements, finite volumes,
finite differences etc. Unfortunately, these computations are
often very expensive and not suitable for online applications such
as analysis, prediction and control. However, solutions of the
original PDE often exhibit low rank representations in an
`optimal' basis~\cite{HLB98}. These representation can be
exploited to reduce the PDE to an ODE of significantly lower
order.

In particular, dimensionality reduction follows three steps: The first
step is to discretize the PDE using a finite number of basis
functions, such as piecewise linear or higher order polynomials or
splines. In this paper we use the well-established finite element
method (FEM), and refer the reader to the large literature, e.g.,
\cite{S97,f83} for details. We denote the approximation of the PDE
solution by $z_n(t,x) \in \mathbb{R}^n$, where $t$ denotes the scalar
time variable, and $x$ denotes the multidimensional space variable,
i.e., $x$ is scalar for a one dimensional space, a vector of two
elements in a two dimensional space, etc. We consider the
one-dimensional case, where $x$ is a scalar in a finite interval,
chosen as $ \Omega = [0,1]$ without loss of generality. By a standard
abuse of notation, $x \in \mathbb{R}^n$ also denotes the
discretization of the spatial domain at equidistant space points, $x_i
= i\cdot \Delta x$, for $i=1,\ldots,n$, and some spatial distance
$\Delta x$. In this notation, $x$ is an $n$-dimensional vector.

The second step is to determine a set of `optimal,' for some
criterion, spatial basis vectors $\phi_i(x) \in \mathbb{R}^n$ for the
discretized problem, that are used to `best' approximate the
discretized PDE solution as
\begin{equation}\label{x_appro1_chap3}
P_n z(t,x) \approx \Phi z_r(t) = \sum_{i=1}^{r} {z_r}_{i}(t) \phi_i(x)
\in \mathbb{R}^n,
\end{equation}
where $P_n$ is the projection of $z(t,x)$ onto $\mathbb{R}^n$. Here,
$\Phi$ is a $n \times r$ matrix containing the basis vectors $\phi_i$
as column vectors. Note that the dimension $n$, coming form the high
fidelity discretization of the PDE described above, is generally very
large, in contrast the dimension $r$ of the optimal basis set.

The third step employs a Galerkin projection, a classical nonlinear
model reduction technique,  to obtain a ROM of the form
\begin{equation}\label{ROM1_chap3}
\dot{z}_r(t) =F(z_r(t)) \in \mathbb{R}^r.
\end{equation}
The function $F:\;\mathbb{R}^r \rightarrow \mathbb{R}^r$ is obtained
from the weak form of the original PDE, through the Galerkin
projection.

The main challenge in this approach lies in the selection of the
`optimal' basis matrix $\Phi$, and the criterion of optimality
used. There are many model reduction methods to find those basis
functions for nonlinear systems. For example some of the most used
methods are proper orthogonal decomposition (POD) \cite{KV02}, dynamic
mode decomposition (DMD) \cite{KGBBN15}, and reduced basis (RB)
\cite{veroy2005certified}.

\begin{remark}
    In the remainder, we present the idea of closure models in the
    framework of POD. However, the derivation it is not limited to a
    particular type of ROM. Indeed, closure model ideas can be applied
    to other ROMs, such as DMD.
\end{remark}

\subsection{Background on POD Basis Functions}
\label{basic_pod_chap3}
We briefly review the necessary steps for computing POD reduced order
models, described in detail in~\cite{KV02,HLB98}. Models based on POD
select basis functions that capture an maximal amount of energy of the
original system. In particular, the POD basis functions are computed
from a collection of snapshots from the dynamical system over a finite
time interval. These snapshots are usually obtained from a discretized
approximation of the PDE model. This approximation could be obtained
using a numerical method, such as FEM, or using direct measurements of
the system modeled by the PDE, if feasible. In this paper, the POD
basis is computed from snapshots of approximate numerical solutions of
the partial differential equation.

To this end, we compute a set of $s$ snapshots of approximate
solutions as
\begin{equation}\label{snap_shot_set_chap3}
\mathcal{S} =\{z_{n}(t_{1},\cdot),...,z_{n}(t_{s},\cdot)\}\subset\mathbb{R}^n,
\end{equation}
where $n$ is the selected number of FEM basis functions, and $t_i$ are
time instances at which snapshots are recorded (do not have to be
uniform). Next, we define the \textit{correlation matrix} $K$ with
elements
\begin{equation}\label{correlation_matrix_pod_chap3}
{K}_{ij}=\frac{1}{s}\langle z_{n}(t_{i},.),z_{n}(t_{j},.)\rangle ,\;i,j=1,...,s.
\end{equation}
The normalized eigenvalues and eigenvectors of $K$ are denoted by
$\lambda_i$ and $v_i$, respectively. Note that the $\lambda_i$ are
also referred to as the \textit{POD eigenvalues}.

The $i^{th}$ \textit{POD basis function} is given by
\begin{equation}\label{pod_basis_chap3}
\phi_{i}(x)=\frac{1}{\sqrt{s}\sqrt{\lambda_{i}}}\sum_{j=1}^{j=s}v_{i,j}z_n (t_{j},x),\;i=1,...,r,
\end{equation}
where $r \leq \min \{s,n \}$, the number of retained POD basis
functions, depends on the application. An important property of the
POD basis functions is their orthonormality:
\begin{equation}\label{pod_ortho_chap3}
\langle \phi_{i},\phi_{j}\rangle =\int_{0}^{1}\phi_{i}(x)\phi_{j}(x)dx=\delta_{ij}
\end{equation}
where $\delta_{ij}$ denotes the Kronecker delta function.

In this new basis, the solution of the PDE (\ref{general_PDE_chap3})
can then be approximated by
\begin{equation}\label{pod_proj_chap3}
z_n^{pod}(t,x)=\sum_{i=1}^{i=r}q_{i}(t) \phi_{i}(x) \ \in \mathbb{R}^n,
\end{equation}
where $q_{i},\;i=1,...,r$ are the POD projection
coefficients (which play the role of the $z_{r,i}(t)$ in the ROM
(\ref{ROM1_chap3})).
To find the coefficients $q_i(t)$, the model (\ref{general_PDE_chap3}) is
projected on the $r^{th}$-order POD space using a Galerkin
projection. In particular, both sides of equation (\ref{general_PDE_chap3})
are multiplied by the POD basis functions, where $z$ is
substituted by $z_n^{pod}$, and then both sides are integrated over
the space interval $\Omega = [0,1]$. Using the orthonormality
of the POD basis (\ref{pod_ortho_chap3}) leads to an ODE of the form
\begin{equation}\label{ROM2_chap3}
\dot{q}(t) =F(q(t)) \in \mathbb{R}^{r}.
\end{equation}
Note, that the Galerkin projection preserves the structure of the nonlinearities of the original PDE.

\subsection{Closure models for ROM stabilization}\label{closure_models_Chap3}  We start with presenting the problem of stable model reduction in its general form, i.e., without specifying a
particular type of PDE. To this end, we highlight the dependence
of the general PDE  (\ref{general_PDE_chap3}), on a single
physical parameter $\mu$ by
\begin{equation}\label{general_PDE2_chap3}
\dot{z} = \mathcal{F}(z,\mu) \in\mathcal{Z},\;\mu\in\mathbb{R}.
\end{equation}
The parameter $\mu$ is assumed to be critical for the stability and
accuracy of the model; changing the parameter can either make the
model unstable, or inaccurate for prediction. As an example, since we
are interested in fluid dynamical problems, we use $\mu$ to denote a
viscosity coefficient. The corresponding reduced order POD model is
takes the form (\ref{pod_proj_chap3}) and (\ref{ROM2_chap3}):
\begin{equation}\label{ROM3_chap3}
\left\{\begin{array}{l} \dot{q}(t) =F(q(t),\mu),\\
z_n^{pod}(t,x)=\sum_{i=1}^{i=r}\phi_{i}(x)q_{i}(t).
\end{array}\right .
\end{equation}
As we explained earlier, the issue with this `simple' Galerkin POD ROM
(denoted POD ROM-G) is that the norm of $z_n^{pod}$ might become
unbounded over a finite time support, despite the fact that the
solution of (\ref{general_PDE2_chap3}) is bounded.

One of the main ideas behind the closure models approach is that the
viscosity coefficient $\mu$ in (\ref{ROM3_chap3}) can be substituted
by a virtual viscosity coefficient $\mu_{cl}$, whose form is chosen to
stabilize the solutions of the POD ROM
(\ref{ROM3_chap3}). Furthermore, a penalty term $H(\cdot,\cdot)$ is
added to the original (POD) ROM-G, as follows
\begin{equation}\label{ROM41_chap3}
\dot{q}(t) =F(q(t),\mu)+H(t,q(t)).
\end{equation}
The term $H(\cdot,\cdot)$ is chosen depending on the structure of
$F(\cdot,\cdot)$ to stabilize the solutions of
(\ref{ROM41_chap3}). For instance, one can use the Cazemier penalty
model described in \cite{SI13}.

\subsection{Closure Model Examples}
The following closure models were introduced in \cite{WABI12,SI13}
for the case of Burgers' equations. We present them in a general
framework, since similar closure terms could be used on other PDE
models. These examples illustrate the principles behind closure
modelling, and motivate our proposed method. Throughout, $r$
denotes the total number of modes retained in the ROM. We recall
below some closure models proposed in the literature, which we
will use later to present the main result of this paper.
\subsubsection{Closure models with constant eddy
viscosity coefficients} Here we describe closure models which are
based on constant stabilizing eddy viscosity coefficients.

  {-\bf\it ROM-H model:} The first eddy
viscosity model, known as the Heisenberg ROM (ROM-H) is simply
given by the constant viscosity coefficient
\begin{equation}\label{PODROMH_CHAP3}
\mu_{cl}=\mu+\mu_{e},
\end{equation}
where $\mu$ is the nominal value of the viscosity coefficient in
(\ref{general_PDE2_chap3}), and $\mu_{e}$ is the additional
constant term added to compensate for the damping effect of the
truncated modes.

{-\bf\it ROM-R model:} This model is a variation of the first one,
introduced in~\cite{R91}. In this model, $\mu_{cl}$ is dependent
on the mode index, and the viscosity coefficients for each mode
are given by
\begin{equation}\label{PODROMR_CHAP3}
\mu_{cl}=\mu+\mu_{e}\frac{i}{r},
\end{equation}
with $\mu_{e}$ being the viscosity amplitude, and $i$ the mode
index.

{-\bf\it ROM-RQ model:} This model proposed in \cite{SI13}, is a
quadratic version of the ROM-R, which we refer to as ROM-RQ. It is
given by the coefficients
\begin{equation}\label{PODROMRQ_CHAP3}
\mu_{cl}=\mu+\mu_{e}\left ( \frac{i}{r} \right ) ^{2},
\end{equation}
where the variables are defined similarly to
(\ref{PODROMR_CHAP3}).

{-\bf\it ROM-RQ model:} This model proposed in \cite{SI13}, is a
root-square version of the ROM-R; we use ROM-RS to refer to it. It
is given by
\begin{equation}\label{PODROMRS_CHAP3}
\mu_{cl}=\mu+\mu_{e}\sqrt{\frac{i}{r}},
\end{equation}
where the coefficients are defined as in (\ref{PODROMR_CHAP3}).

{-\bf\it ROM-T model:} Known as spectral vanishing viscosity
model, is similar to the ROM-R in the sense that the amount of
induced damping changes as function of the mode index. This
concept has been introduced by Tadmor in \cite{T89}, and so these
closure models are referred to as ROM-T. These models are given by
\begin{equation}\label{PODROMRT_CHAP3}
\left\{\begin{array}{l} \mu_{cl}=\mu,\;\;\;\;\;\;\;\;\;\text{for} \;i \leq m\\
\mu_{cl}=\mu+\mu_{e},\;\text{for}\;i>  m\\
\end{array}\right.
\end{equation}
where $i$ denotes the mode index, and $m\leq r$ is the index of
modes above which a nonzero damping is introduced.

{-\bf\it ROM-SK model:} Introduced by Sirisup and Karniadakis in
\cite{SK04}, falls into the class of vanishing viscosity models.
We use ROM-SK to refer to it; it is given by
\begin{equation}\label{PODROMSK_CHAP3}
\left\{\begin{array}{l} \mu_{cl}=\mu+\mu_{e}e^{\frac{-(i-r)^{2}}{(i-m)^{2}}},\;\text{for} \;i \leq m\\
\mu_{cl}=\mu,\;\;\;\;\;\;\;\;\;\;\;\text{for}\;i>m,\;m\leq r\\
\end{array}\right.
\end{equation}

{-\bf\it ROM-CLM model:} This model has been introduced in
\cite{C84,LM96}, and is given by
\begin{equation}\label{PODROMCLM_CHAP3}
\mu_{cl}=\mu+\mu_{e}\alpha_{0}^{-1.5}(\alpha_{1}+\alpha_{2}e^{-\frac{\alpha_{3}r}{i}}),
\end{equation}
where $i$ is the mode index, and
$\alpha_{0},\;\alpha_{1},\;\alpha_{2},\;\alpha_{3}$ are positive
gains (see \cite{KK00,C84} for some insight about their tuning).

\subsubsection{Closure models with time and space varying eddy
viscosity coefficients}

Several (time and/or space) varying viscosity terms have been proposed
in the literature. For instance, \cite{SI13} describes the Smagorinsky
nonlinear viscosity model. However, the model requires online
computation of some nonlinear closure terms at each time step, which
in general makes it computationally consuming. We report here the
nonlinear viscosity model presented in \cite{CNTLDBDN13}, which is
nonlinear and a function of the ROM state variables. This requires
explicit rewriting of the ROM model (\ref{ROM3_chap3}), to separate
the linear viscous term as follows
\begin{equation}\label{ROM4_chap3}
\left\{\begin{array}{l}
\dot{q}(t) =F(q(t),\mu)=\tilde{F}(q(t))+\mu\;Dq(t)\\
z_n^{pod}(t,x)=\sum_{i=1}^{i=r}\phi_{i}(x)q_{i}(t),
\end{array}\right.
\end{equation}
where $D\in\mathbb{R}^{r\times r}$ represents a constant viscosity
damping matrix, and the term $\tilde{F}$ represents the remainder
of the ROM model, i.e., the part without damping. \\Based on
equation (\ref{ROM4_chap3}), we can write the nonlinear eddy
viscosity model denoted by $H_{nev}(.)$, as
\begin{equation}\label{Smagorinsky_CHAP3}
H_{nev}(\mu_e,
q(t))=\mu_{e}\sqrt{\frac{V(q(t))}{V_{\infty}(\lambda)}}diag(d_{11},...,d_{rr})q(t),
\end{equation}
where $\mu_{e}>0$ is the amplitude of the closure model, the
$d_{ii},\;i=1,\ldots,r$ are the diagonal elements of the matrix
$D$, and $V(q),\;V_{\infty}(\lambda)$ are defined as follows
\begin{equation}\label{V1_CHAP3}
V(q)=\frac{1}{2}\sum_{i=1}^{i=r}{q_{i}}^{2},
\end{equation}
\begin{equation}\label{Vinfty1_CHAP3}
V_{\infty}(\lambda)=\frac{1}{2}\sum_{i=1}^{i=r}\lambda_{i},
\end{equation}
where the $\lambda_{i}$ are the selected POD eigenvalues (as
defined in Section \ref{basic_pod_chap3}). Compared to the
previous closure models, the nonlinear term $H_{nev}$ does not
just act as a viscosity, but is rather added directly to the
right-hand side of the reduced order model (\ref{ROM4_chap3}), as
an additive stabilizing nonlinear term. The stabilizing effect has
been analyzed in \cite{CNTLDBDN13} based on the decrease over time
of an energy function along the trajectories of the ROM solutions,
i.e., a Lyapunov-type analysis.

All these closure models share several characteristics, including a
common challenge, among others~\cite{CNTLDBDN13,SB14}: the selection
and tuning of their free parameters, such as the closure models
amplitude $\mu_{e}$. In the next section, we show how ES can be used
to auto-tune the closure models' free coefficients and optimize their
stabilizing effect.

\subsection{Main result: ES-based closure models auto-tuning}
\label{MES-closure-tuning_chap3}
As mentioned in~\cite{SB14}, the tuning of the closure model amplitude
is important to achieve an optimal stabilization of the ROM. To
achieve optimal stabilization, we use model-free ES optimization
algorithms to tune the coefficients of the closure models presented in
Section \ref{closure_models_Chap3}. The advantage of using ES is the
auto-tuning capability that such algorithms allow. Moreover, in
contrast to manual off-line tuning approaches, the use of ES allows us
to constantly tune the closure model, even in an online operation of
the system. Indeed, ES can be used off-line to tune the closure model,
but it can also be connected online to the real system to continuously
fine-tune the closure model coefficients, such as the amplitudes of
the closure models. Thus, the closure model can be valid for a longer
time interval compared to the classical closure models with constant
coefficients, which are usually tuned off-line over a fixed finite
time interval.

We start by defining a suitable learning cost function. The goal of
the learning (or tuning) is to enforce Lagrange stability of the ROM
model (\ref{ROM3_chap3}), and to ensure that the solutions of the ROM
(\ref{ROM3_chap3}) are close to the ones of the original PDE
(\ref{general_PDE2_chap3}). The later learning goal is important for
the accuracy of the solution. Model reduction works toward obtaining a
simplified ODE model which reproduces the solutions of the original
PDE (the real system) with much less computational burden, i.e., using
the lowest possible number of modes. However, for model reduction to
be useful, the solution should be accurate.

We define the learning cost as a positive definite function of the
norm of the error between the approximate solutions of
(\ref{general_PDE2_chap3}) and the ROM (\ref{ROM3_chap3}), as follows
\begin{equation}\label{Q_pde2_chap3}
\begin{array}{l}
Q(\hat\mu)=\tilde{H}(e_{z}(t,\hat \mu)),\\
e_{z}(t,\hat{\mu})=z_n^{pod}(t,x,\hat{\mu})-z_n(t,x,\mu),
\end{array}
\end{equation}
where $\hat\mu\in\mathbb{R}$ denotes the learned parameter, and
$\tilde{H}$ is a positive definite function of $e_{z}$. Note that the
error $e_{z}$ could be computed off-line using solutions of the ROM
(\ref{ROM3_chap3}), and approximate solutions of the PDE
(\ref{general_PDE2_chap3}). The error could be also computed online
where the $z_n^{pod}(t,x,\hat{\mu})$ is obtained from solving the
model (\ref{ROM3_chap3}), but the $z_n(t,x,\mu)$ is obtained from real
measurements of the system at selected space points $x$.

A more practical way of implementing the ES-based tuning of
$\hat\mu$, is to start with an off-line tuning of the closure
model. Then, the obtained ROM, i.e., the computed optimal value of
$\hat\mu$, is used in an online operation of the system, e.g.,
control and estimation. One can then fine-tune the ROM online by
continuously learning the best value of $\hat\mu$ at any give time
during the operation of the system.

To derive formal convergence results, we use some classical
assumptions of the solutions of the original PDE, and on the learning
cost function.
\begin{assumption}\label{pdestab_assumption1_chap3}
The solutions of the original PDE model
(\ref{general_PDE2_chap3}), are assumed to be in
$L^{2}([0,\infty);\mathcal{Z})$, $\forall\mu\in\mathbb{R}$.
\end{assumption}
\begin{assumption} \label{robustmesass1_pdestab_chap3}
The cost function $Q$ in (\ref{Q_pde2_chap3}) has a local minimum
at $\hat\mu= \mu^{*}$.
\end{assumption}
\begin{assumption} \label{robustmesass2_pdestab_chap3}
The cost function $Q$ in (\ref{Q_pde2_chap3}) is analytic and its
variation with respect to $\mu$ is bounded in the neighborhood of
$\mu^{*}$, i.e., $\|\frac{\partial{Q}}{\partial
\mu}({\tilde{\mu}})\|\leq\xi_{2},\;\xi_{2}>0,
\;\tilde{\mu}\in\mathcal{V}(\mu^{*})$, where
$\mathcal{V}(\mu^{*})$ denotes a compact neighborhood of
$\mu^{*}$.
\end{assumption}

Under these assumptions, the following lemma follows.
\begin{lemma}\label{pdestab_lemma1_chap3}
Consider the PDE (\ref{general_PDE2_chap3}), under Assumption
\ref{pdestab_assumption1_chap3}, together with its ROM model
(\ref{ROM3_chap3}), where the viscosity coefficient $\mu$ is
substituted by $\mu_{cl}$. Let $\mu_{cl}$ take the form of any of
the closure models in (\ref{PODROMH_CHAP3}) to
(\ref{PODROMCLM_CHAP3}), where the closure model amplitude
$\mu_{e}$ is tuned based on the following ES algorithm
\begin{equation}
\begin{array}{l}
\dot{y}=a\;\sin(\omega t+\frac{\pi}{2})Q(\hat\mu_{e}),\\
\hat{\mu}_{e}=y+a\;\sin(\omega t-\frac{\pi}{2}),
\label{pdestab_mes_1_chap3}
\end{array}
\end{equation}
where $y(0)=0$, $\omega>\omega^{*}$, $\omega^{*}$ large enough,
and $Q$ is given by (\ref{Q_pde2_chap3}). Under Assumptions
\ref{robustmesass1_pdestab_chap3}, and
\ref{robustmesass2_pdestab_chap3}, the norm of the distance w.r.t.
the optimal value of $\mu_{e}$,
 $e_{\mu}=\mu^{*}-{\hat\mu}_{e}(t)$ admits the following bound
\begin{equation}
\|e_{\mu}(t)\|\leq\frac{\xi_{1}}{\omega}+a,\;t\rightarrow\infty
\label{pdestab_mes_bound1_chap3}
\end{equation}
where $a>0,\;\xi_{1}>0$, and the learning cost function approaches
its optimal value within the following upper-bound\
\begin{equation}
\begin{array}{l}
\|Q({\hat\mu}_{e})-Q(\mu^{*})\|\leq\xi_{2}(\frac{\xi_{1}}{\omega}+a),\;t\rightarrow\infty
\end{array}\label{pdestab_mes_bound2_chap3}
\end{equation}
where $\xi_{2}=\underset{\mu\in\mathcal{V}(\mu^{*})}\max
\|\frac{\partial{Q}}{\partial\mu} \|$.
\end{lemma}
\begin{pf}
Based on Assumptions \ref{robustmesass1_pdestab_chap3}, and
\ref{robustmesass2_pdestab_chap3}, the extremum seeking nonlinear
dynamics (\ref{pdestab_mes_1_chap3}), can be approximated by
linear averaged dynamics (using averaging approximation over time,
\cite[p. 435, Definition 1]{Rote00}). Furthermore, there exist $
\xi_{1},\;\omega^{*}$, such that for all $\omega>\omega^{*}$, the
solution of the averaged model ${\hat\mu}_{av}(t)$ is locally
close to the solution of the original ES dynamics, and satisfies
(\cite[p. 436]{Rote00})
$$
\|{\hat\mu}_{e}(t)-d(t)-{\hat\mu}_{av}(t)\|
\leq\frac{\xi_{1}}{\omega},\;\xi_{1}>0,\;\forall t\geq 0,
$$
with $d(t)=a \sin(\omega t-\frac{\pi}{2})$. Moreover, since $Q$ is
analytic it can be approximated locally in $\mathcal{V}(\mu^{*})$
with a quadratic polynomial, e.g., Taylor series up to second
order, which leads to (\cite[p. 437]{Rote00})
$$
\lim_{t\rightarrow\infty}{\hat\mu}_{av}(t)=\mu^{*}.
$$
Based on the above, we can write
$$
\begin{array}{c}\|{\hat\mu}_{e}(t)-\mu^{*}\|-\|d(t)\|\leq |{\hat\mu}_{e}(t)-\mu^{*}-d(t)\|
\leq\frac{\xi_{1}}{\omega},\;\xi_{1}>0,\\\hspace{+7cm}t\rightarrow\infty\\
\Rightarrow \|{\hat\mu}_{e}(t)-\mu^{*}\|
\leq\frac{\xi_{1}}{\omega}+\|d(t)\|,\;t\rightarrow\infty,
\end{array}
$$
which implies
$$
\begin{array}{c}
\|{\hat\mu}_{e}(t)-\mu^{*}\|\leq
\frac{\xi_{1}}{\omega}+a,\;\xi_{1}>0,\;t\rightarrow\infty.
\end{array}
$$
The cost function upper-bound is easily obtained from the previous
bound, using the fact that $Q$ is locally Lipschitz, with the
Lipschitz constant
$\xi_{2}=\underset{\mu\in\mathcal{V}(\mu^{*})}\max \|
\frac{\partial{Q}}{\partial\mu} \|$.
 \hspace{+9cm}$\Box$
\end{pf}

When the influence of the linear terms of the PDE are dominant, e.g.,
in short-time scales, closure models based on constant linear eddy
viscosity coefficients can be a good solution to stabilize ROMs and
preserve the intrinsic energy properties of the original PDE. However,
in many cases with nonlinear energy cascade, these closure models are
unrealistic; linear terms cannot recover the nonlinear energy terms
lost during the ROM computation. For this reason, many researchers
have tried to come up with nonlinear stabilizing terms for instable
ROMs. An example of such a nonlinear closure model is the one given by
equation (\ref{Smagorinsky_CHAP3}), and proposed in \cite{NETAL08}
based on finite-time thermodynamics (FTT) arguments and in
\cite{NMT011} based on scaling arguments.

Based on the above, we introduce here a combination of both linear and
nonlinear closure models. The combination of both models can lead to a
more efficient closure model. In particular, this combination can
efficiently handle linear energy terms, that are typically dominant
for small time scales and handle nonlinear energy terms, which are
typically more dominant for large time-scales and in some specific
PDEs/boundary conditions. Furthermore, we propose to auto-tune this
closure model using ES algorithms, which provides an automatic way to
select the appropriate term to amplify. It can be either the linear
part or the nonlinear part of the closure model, depending on the
present behavior of the system, e.g., depending on the test
conditions.  We summarize this result in the following Lemma.

\begin{lemma}\label{pdestab_lemma2_chap3}
Consider the PDE (\ref{general_PDE2_chap3}), under Assumption
\ref{pdestab_assumption1_chap3}, together with its stabilized ROM
model
\begin{equation}\label{pdestab_romnl1_CHAP3}
\left\{\begin{array}{l}
\dot{q}(t) =F(q(t),\mu)=\tilde{F}(q(t))+\mu_{lin}\;Dq(t)+H_{nl}(q,\mu_{nl})\\
z_n^{pod}(t,x)=\sum_{i=1}^{i=r}\phi_{i}(x)q_{i}(t)\\
H_{nl}=\mu_{nl}\sqrt{\frac{V(q)}{V_{\infty}(\lambda)}}diag(d_{11},...,d_{rr})q(t)\\
V(q)=\frac{1}{2}\sum_{i=1}^{i=r}{q_{i}}^{2}\\
V_{\infty}(\lambda)=\frac{1}{2}\sum_{i=1}^{i=r}\lambda_{i},
\end{array}\right.
\end{equation}
 where the linear viscosity coefficient
$\mu_{lin}$ is substituted by $\mu_{cl}$ chosen from any of the
constant closure models (\ref{PODROMH_CHAP3}) to
(\ref{PODROMCLM_CHAP3}). The closure model amplitudes
$\mu_{e},\;\mu_{nl}$ are tuned based on the following ES algorithm
\begin{equation}
\begin{array}{l}
\dot{y}_{1}=a_{1}\;\sin(\omega_{1} t+\frac{\pi}{2})Q(\hat\mu_{e},{\hat\mu}_{nl}),\\
\hat{\mu}_{e}=y_{1}+a_{1}\;\sin(\omega_{1} t-\frac{\pi}{2}),\\
\dot{y}_{2}=a_{2}\;\sin(\omega_{2} t+\frac{\pi}{2})Q(\hat\mu_{e},{\hat\mu}_{nl}),\\
\hat{\mu}_{nl}=y_{2}+a_{2}\;\sin(\omega_{2} t-\frac{\pi}{2}),
\label{pdestab_mes_2_chap3}
\end{array}
\end{equation}
where $y_{1}(0)=y_{2}(0)=0$,
$\omega_{max}=max\{\omega_{1},\omega_{2}\}>\omega^{*}$,
$\omega^{*}$ large enough, and $Q$ is given by
(\ref{Q_pde2_chap3}), with
$\hat\mu=({\hat\mu}_{e},{\hat\mu}_{nl})$. Under Assumptions
\ref{robustmesass1_pdestab_chap3}, and
\ref{robustmesass2_pdestab_chap3}, the norm of the vector of
distance w.r.t. the optimal values of $\mu_{e},\;\mu_{nl}$;
 $e_{\mu}=({\mu_{e}}^{*}-{\hat\mu}_{e}(t),{\mu_{nl}}^{*}-{\hat\mu}_{nl}(t))$ admits the following bound
\begin{equation}
\begin{array}{l}
\|e_{\mu}(t)\|\leq\frac{\xi_{1}}{\omega_{max}}+\sqrt{a_{1}^{2}+a_{2}^{2}},\;t\rightarrow\infty
\end{array}\label{pdestab_mes_bound3_chap3}
\end{equation}
where $a_{1},\;a_{2}>0,\;\xi_{1}>0$, and the learning cost
function approaches its optimal value within the following
upper-bound
\begin{equation}
\begin{array}{l}
\|Q({\hat\mu}_{e},{\hat\mu}_{nl})-Q({\mu_{e}}^{*},{\mu_{nl}}^{*})\|\leq\xi_{2}(\frac{\xi_{1}}{\omega}+\sqrt{a_{1}^{2}+a_{2}^{2}}),\\\hspace{+7cm}t\rightarrow\infty
\end{array}\label{pdestab_mes_bound4_chap3}
\end{equation}
where
$\xi_{2}=\underset{(\mu_{1},\mu_{2})\in\mathcal{V}(\mu^{*})}\max
 \| \frac{\partial{Q}}{\partial\mu} \|$.
\end{lemma}
\begin{pf}
We will skip the proof for this Lemma, since it follows the same
steps as the proof of Lemma \ref{pdestab_lemma1_chap3}.
\end{pf}

\section{Example: The coupled Burgers' equation}
\label{Burgers-example}
As an example application of our approach, we consider the coupled
Burgers' equation, (e.g., see \cite{B_master_011}), of the form
\begin{equation}\label{Burgers2_chap3}
\left\{\begin{array}{l} \frac{\partial w(t,x)}{\partial
t}+w(t,x)\frac{w(t,x)}{\partial
x}=\mu\frac{\partial^{2}w(t,x)}{{\partial x}^{2}}-\kappa
T(t,x)\\
\frac{\partial T(t,x)}{\partial t}+w(t,x)\frac{\partial
T(t,x)}{\partial x}=c\frac{\partial^{2}T(t,x)}{{\partial
x}^{2}}+f(t,x),
\end{array}\right.
\end{equation}
where $T(\cdot,\cdot)$ represents the temperature,
$w(\cdot,\cdot)$ represents the velocity field, $\kappa$ is the
coefficient of the thermal expansion, $c$ the heat diffusion
coefficient, $\mu$ the viscosity (inverse of the Reynolds number
$Re$), $x\in[0,1]$ is the one dimensional space variable, $t>0$,
and $f(\cdot,\cdot)$ is the external forcing term such that $f\in
L^{2}((0,\infty),X),\;X=L^{2}([0,1])$. The boundary conditions are
imposed as
\begin{equation}\label{boundarycd1_chap3}
\begin{array}{l} w(t,0)=w_L,\;\;\frac{\partial w(t,1)}{\partial x}=w_R,\\
T(t,0)=T_L,\;\;\;T(t,1)=T_R,
\end{array}
\end{equation}
where $w_L, w_R, T_L,T_R$ are positive constants, and $L$ and $R$
denote left and right boundary, respectively. The initial conditions
are imposed as
\begin{equation}\label{initialcd1_chap3}
\begin{array}{l}
w(0,x)=w_{0}(x)\in L^{2}([0,1]),\\
T(0,x)=T_{0}(x)\in L^{2}([0,1]),
\end{array}
\end{equation}
and are specified below. Following a Galerkin projection onto the
subspace spanned by the POD basis functions, the coupled Burgers'
equation is reduced to a POD ROM with the following structure (e.g.,
see \cite{B_master_011})
\begin{equation}\label{Bur-PODROM2-CHAP3}
\begin{array}{l}
\left(\begin{array}{l}
{{\dot{q}}_{w}}\\{{\dot{q}}_{T}}
\end{array}\right)
=B_{1}+\mu B_{2}+\mu\;D\;q+\tilde{D}q+Cq{q}^{T},\\
w^{pod}_{n}(x,t)=w_{av}(x)+\sum_{i=1}^{i=r}\phi_{wi}(x)q_{w i}(t),\\
T^{pod}_{n}(x,t)=T_{av}(x)+\sum_{i=1}^{i=r}\phi_{Ti}(x)q_{T i}(t),
\end{array}
\end{equation}
where matrix $B_{1}$ is due to the projection of the forcing term
$f$, matrix $B_{2}$ is due to the projection of the boundary
conditions, matrix $D$ is due to the projection of the viscosity
damping term $\mu\frac{\partial^{2}w(t,x)}{{\partial x}^{2}}$,
matrix $\tilde{D}$ is due to the projection of the thermal
coupling and the heat diffusion terms $-\kappa
T(t,x),\;c\frac{\partial^{2}T(t,x)}{{\partial x}^{2}}$, and the
matrix $C$ is due to the projection of the gradient-based terms
$w\frac{w(t,x)}{\partial x}$, and $w\frac{\partial
T(t,x)}{\partial x}$. The notations $\phi_{w i}(x),\;q_{w i}(t)$
($i=1,...,r_w$), $\phi_{T i}(x),\;q_{T i}(t)$ ($i=1,...,r_T$),
stand for the space basis functions and the time projection
coordinates, for the velocity and the temperature, respectively.
The terms $w_{av}(x),\;T_{av}(x)$ represent the mean values (over
time) of $w$ and $T$, respectively.

\subsection{Burgers equation
ES-based POD ROM stabilization} \subsubsection{Test 1} 
\begin{figure}\center
  \begin{minipage}{0.4\linewidth}
 \center\subfigure[True velocity profile]{
\includegraphics[width=1\linewidth]{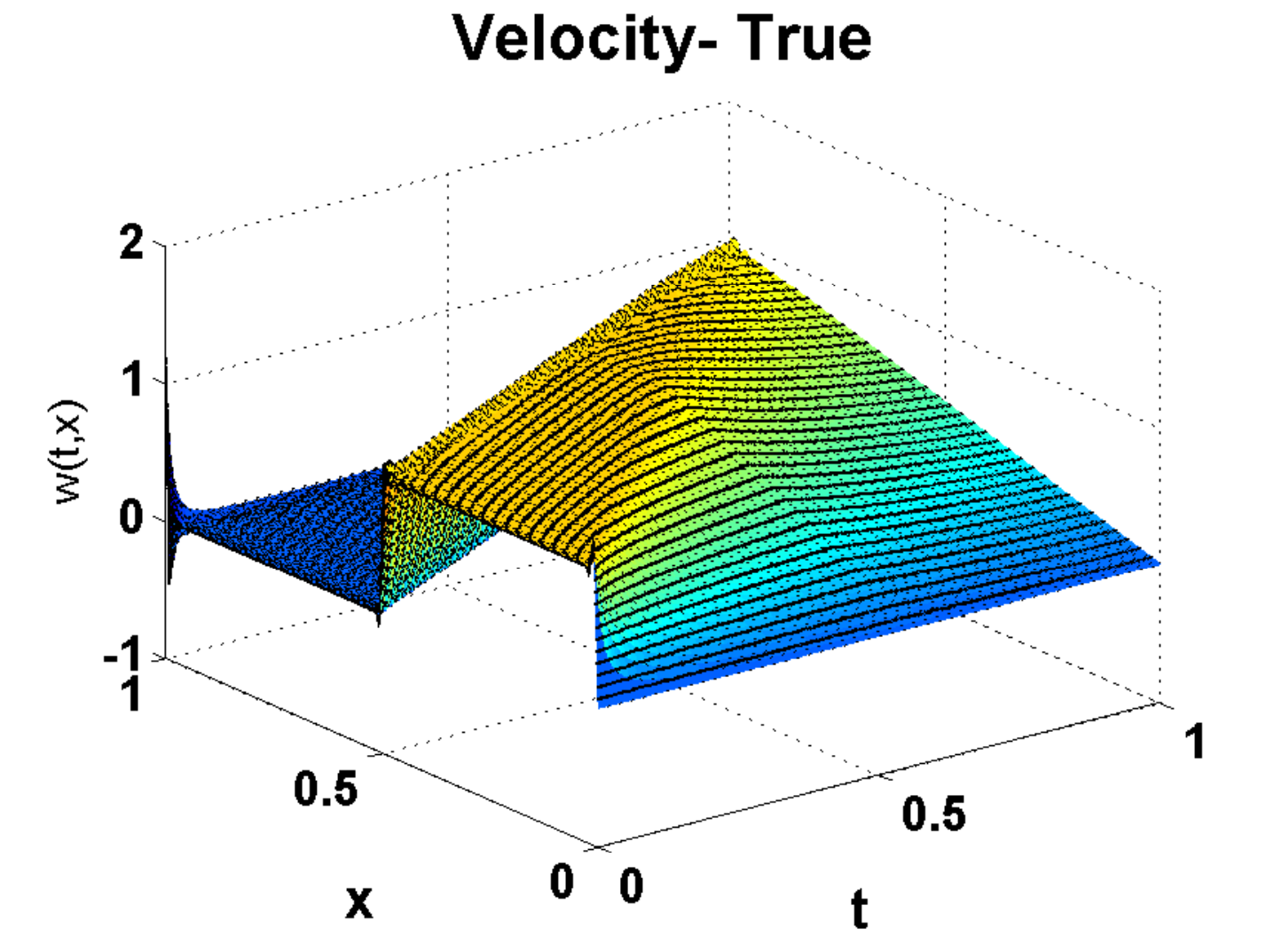}
\label{burgersstab_truev_test1_chap3}}
  \end{minipage}
  \hfill
  \begin{minipage}{0.4\linewidth}
   \center\subfigure[True temperature profile]{
\includegraphics[width=1\linewidth]{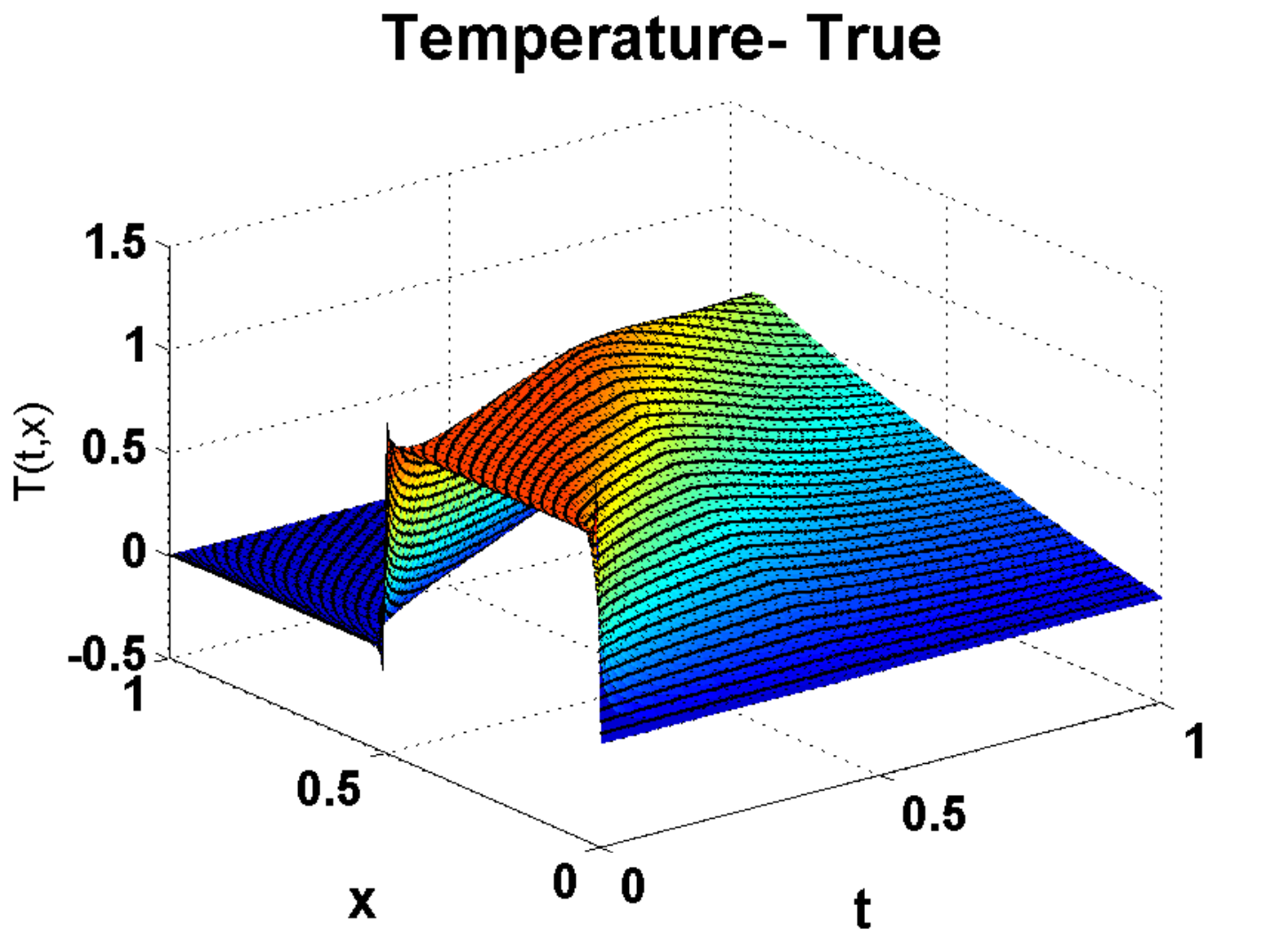}
\label{burgersstab_truet_test1_chap3}}
  \end{minipage}
  \caption{True solutions of (\ref{Burgers2_chap3})}
  \label{burgersstab_truesol_test1_chap3}
\end{figure}

\begin{figure}\center
  \begin{minipage}{0.4\linewidth}
   \center\subfigure[Learning cost function vs. number of iterations]{
\includegraphics[width=1\linewidth]{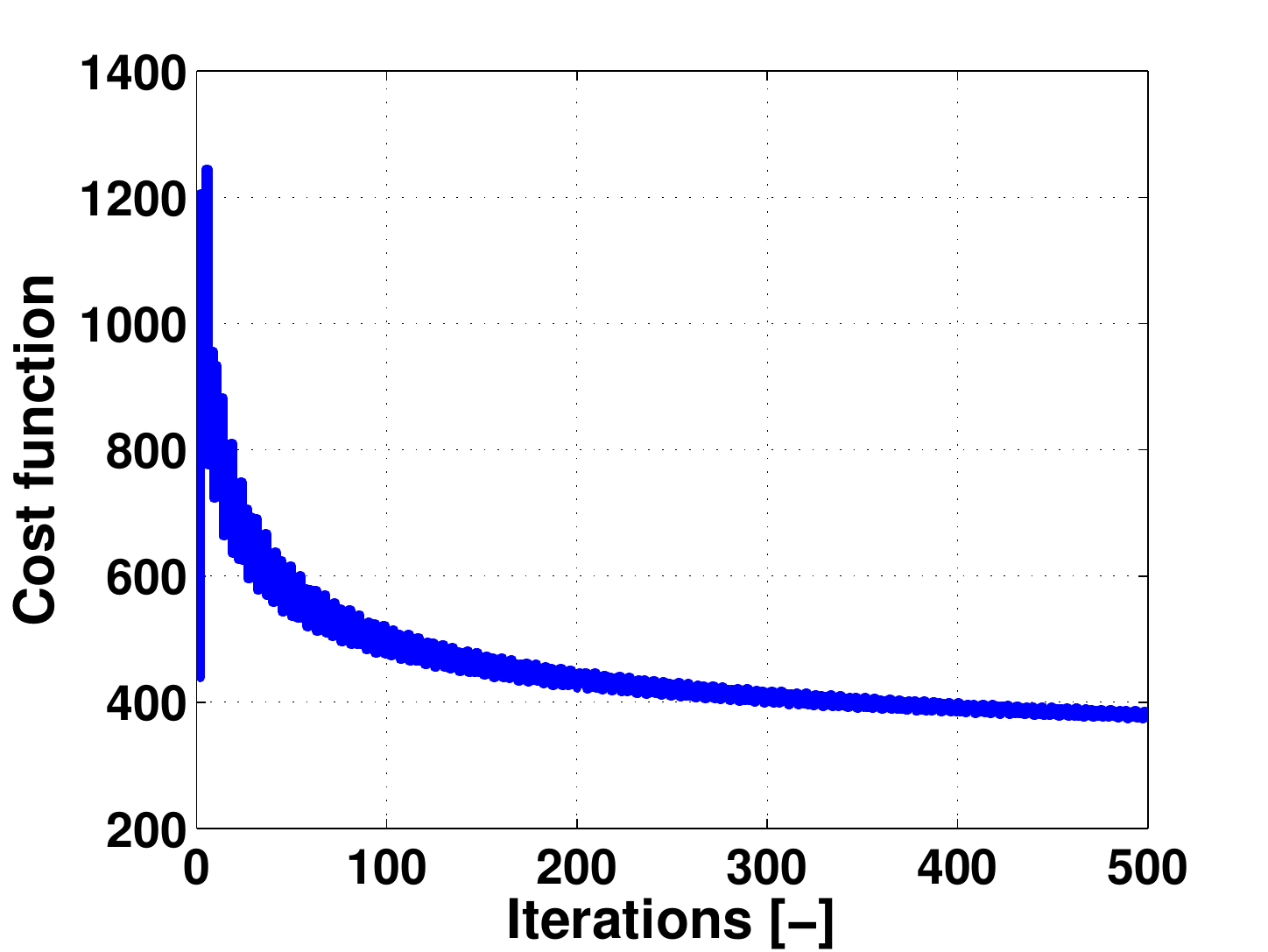}
\label{burgersstab_Q_test1lin_chap3}}
  \end{minipage}
  \hfill
  \begin{minipage}{0.4\linewidth}
   \center\subfigure[Learned parameter $\hat{\nu_{e}}$ vs. number of iterations]{
\includegraphics[width=1\linewidth]{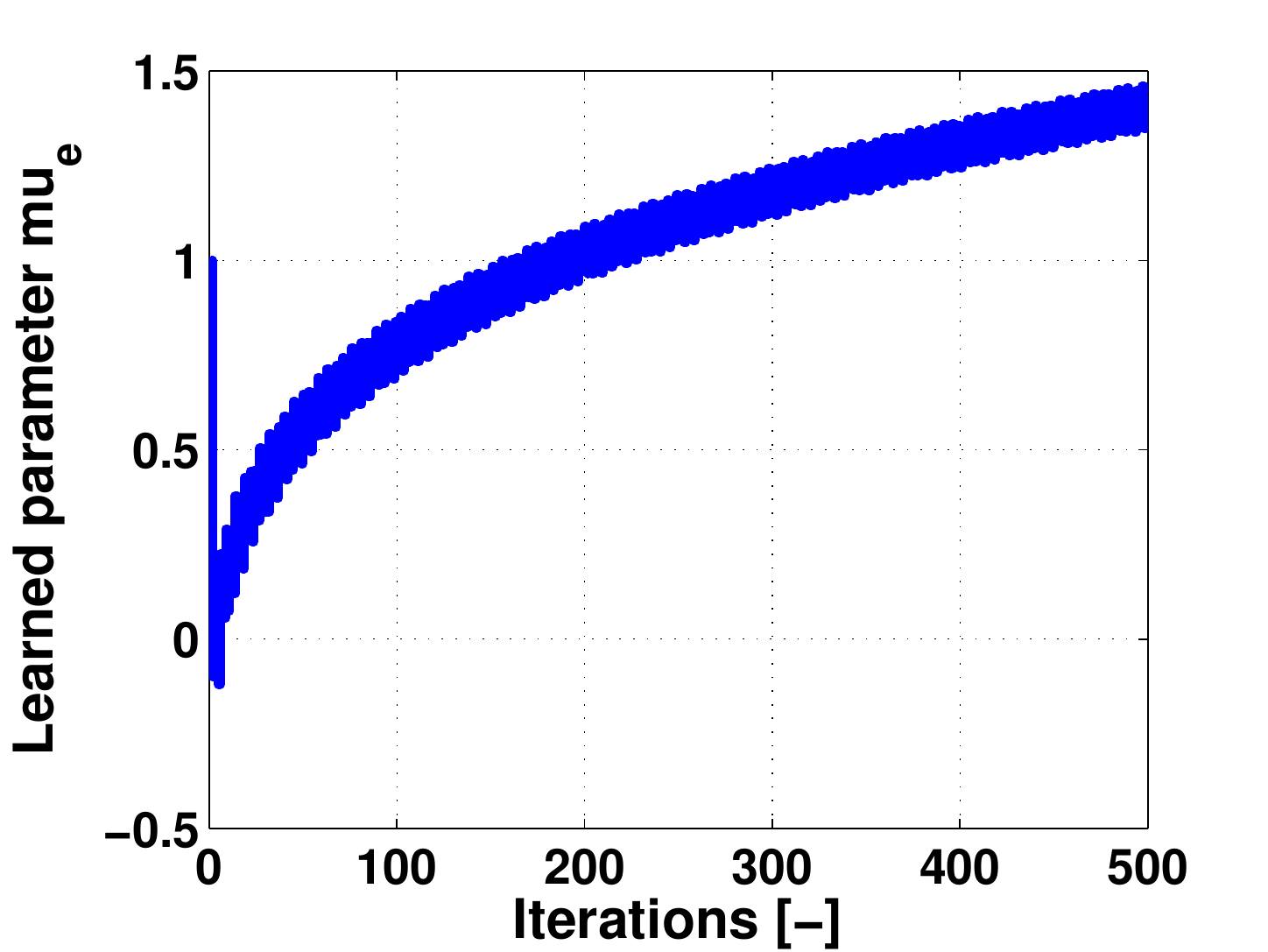}
\label{burgersstab_hatnue_test1lin_chap3}}
  \end{minipage}
  \caption{Learned parameters and learning cost function- Stabilization with learning- Linear closure model}
  \label{burgersstab_learningpara_test1lin_chap3}
\end{figure}

\begin{figure}\center
  \begin{minipage}{0.4\linewidth}
   \center\subfigure[Learning-based POD ROM velocity profile]{
\includegraphics[width=1\linewidth]{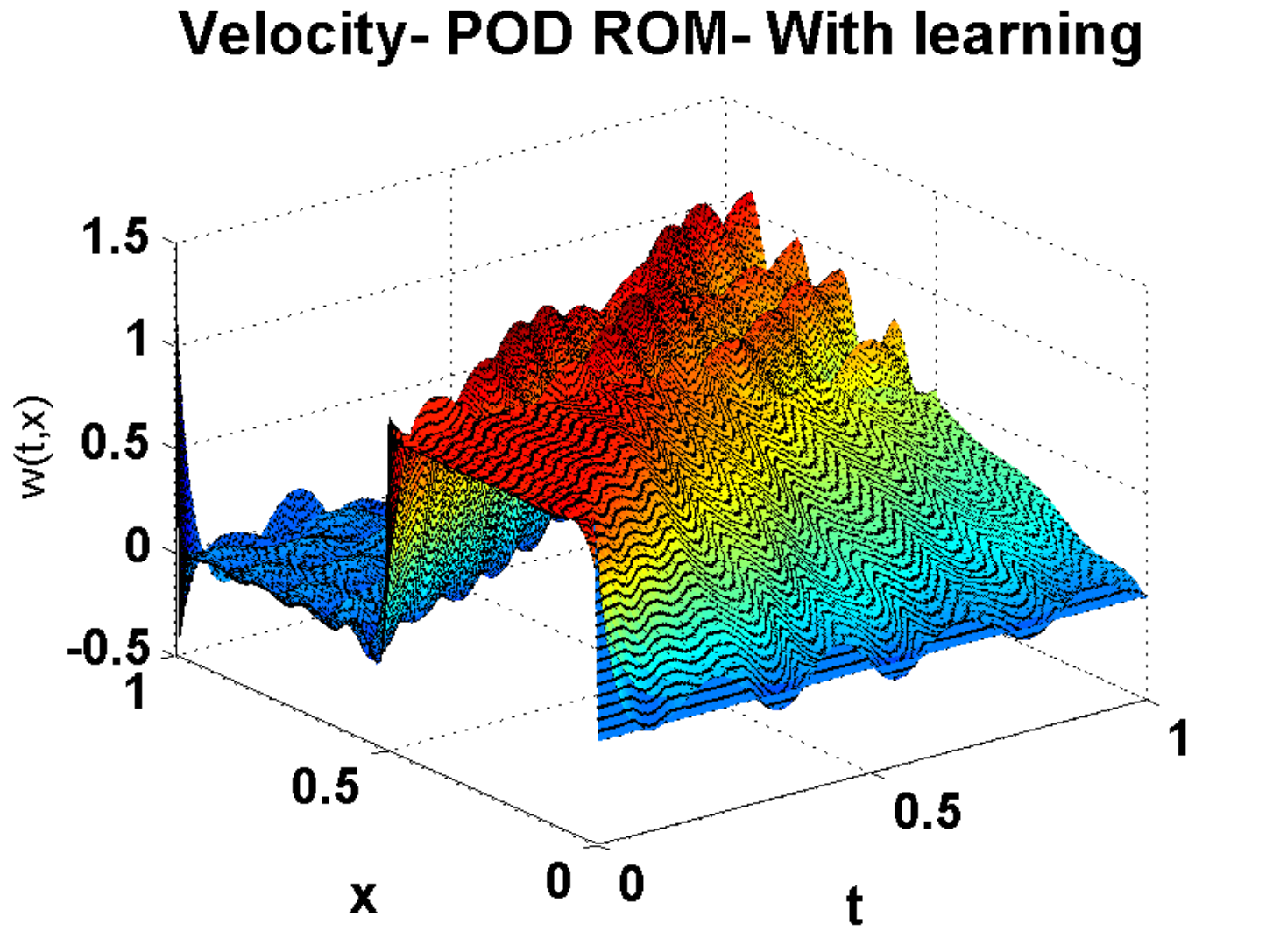}
\label{burgersstab_lromv_test1lin_chap3}}
  \end{minipage}
  \hfill
  \begin{minipage}{0.4\linewidth}
   \center\subfigure[Learning-based POD ROM temperature profile]{
\includegraphics[width=1\linewidth]{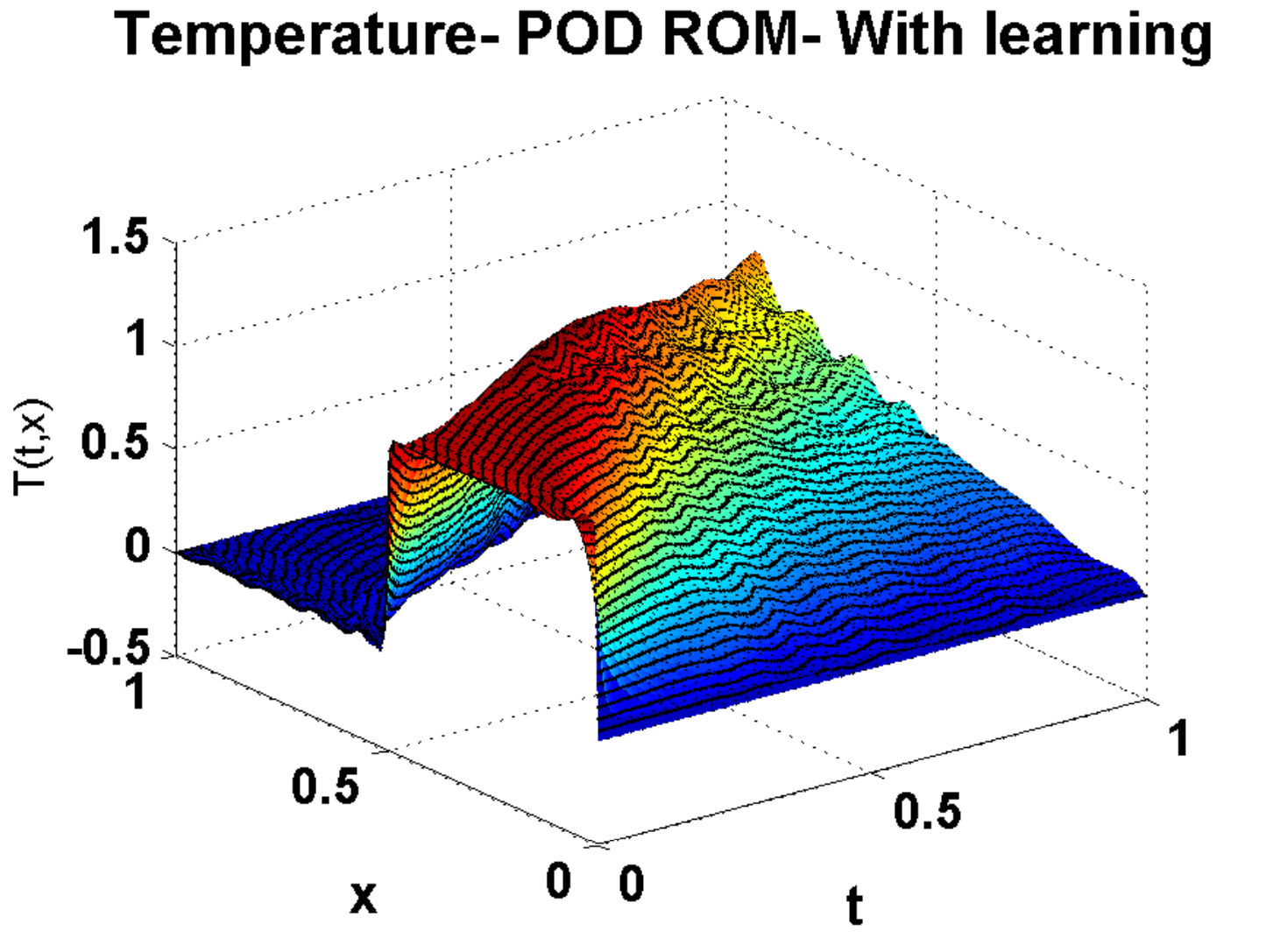}
\label{burgersstab_lromt_test1lin_chap3}}
  \end{minipage}
  \caption{Learning-based POD ROM solutions of (\ref{Burgers2_chap3})- Stabilization with learning- Linear closure model}
  \label{burgersstab_lromsol_test1lin_chap3}
\end{figure}

We first
test the stabilization performance of Lemma
\ref{pdestab_lemma1_chap3}. We test the auto-tuning results of the
ES learning algorithm when tuning the amplitudes of linear closure
models. More specifically, we test the performance of the
Heisenberg closure model given by (\ref{PODROMH_CHAP3}), when
applied in the context of Lemma \ref{pdestab_lemma1_chap3}.
 We consider the coupled Burgers'
equation (\ref{Burgers2_chap3}), with the parameters
$Re=1000,\;\kappa=5\times 10^{-4},\;c=1\times 10^{-2}$,  the
trivial boundary conditions $w_L=w_R=0, \ T_L=T_R=0$, a simulation
time-length $t_{f}=1s$ and zero forcing,  $f =0$. We use $10$ POD
modes for both variables (temperature and velocity). For the
choice of the initial conditions, we follow \cite{SI13}, where the
simplified Burgers' equation has been used in the context of POD
ROM stabilization. Indeed, in \cite{SI13} the authors propose two
types of initial conditions for the velocity variable, which led
to instability of the nominal POD ROM, i.e., the basic Galerkin
POD ROM (POD ROM-G) without any closure model. Accordingly, we
choose the following initial conditions:
\begin{equation}\label{vinitial_cond_burgers_stab_test1}
w(x,0)= \left\{
\begin{array}{l}
1,\;\text{if}\;x\in\;[0,\;0.5]\\
0,\;\text{if}\;x\in\;]0.5,\;1],
\end{array}
\right.
\end{equation}
\begin{equation}\label{Tinitial_cond_burgers_stab_test1}
T(x,0)= \left\{
\begin{array}{l}
1,\;\text {if}\;x\in\;[0,\;0.5]\\
0,\;\text {if}\;x\in\;]0.5,\;1],
\end{array}
\right.
\end{equation}
We apply Lemma \ref{pdestab_lemma1_chap3}, with the Heisenberg
linear closure model given by (\ref{PODROMH_CHAP3}). The closure
model amplitude $\mu_{e}$ is tuned using a discretized version of
the ES algorithm (\ref{pdestab_mes_1_chap3}), given by
\begin{equation}
\begin{array}{l}
y(k+1)=y(k)+a\;t_{f}\;\sin(\omega t_{f}k+\frac{\pi}{2})Q(\hat\mu_{e}), a>0,\\
\hat{\mu}_{e}(k+1)=y(k+1)+a\;\sin(\omega t_{f}k-\frac{\pi}{2}),
\label{pdestab_mes_1discrt_chap3}
\end{array}
\end{equation}
 where $y(0)=0$, and $k=0,1,2,...,$ denotes the learning iterations index. We use the
 parameters' values: $a_{1}=3\;10^{-4}\;[-],\;\omega_{1}=15\;[\frac{rad}{sec}]$. The learning
cost function is chosen as
\begin{equation}\label{Re_estim_Q1_chap3_}
Q=Q_1\int_{0}^{t_{f}}<e_{T},e_{T}>_{X}dt+Q_2\int_{0}^{t_{f}}<e_{w},e_{w}>_{X}dt,
\end{equation}
with $Q_{1}=Q_{2}=1$ for this example. Moreover,
$e_{T}=T_{n}-T^{pod}_{n},\;e_{w}=w_{n}-w^{pod}_{n}$ define the
errors between the true model solution and the POD ROM solution
for temperature and velocity, respectively.\\We first show for
comparison purposes the true solutions, the high fidelity
(`truth') solutions obtained via the FEM. Equation
(\ref{Burgers2_chap3}) is discretized in space with piecewise
linear basis functions with 100 elements, and subsequently solved
with \texttt{ode45} in \texttt{Matlab}\textregistered. The true
solutions are reported in Figure
\ref{burgersstab_truesol_test1_chap3}. We then report in figure
\ref{burgersstab_lfromsol_test1_chap3}, the solutions of the POD
ROM-G (without learning). We can see clearly in these figures that
the POD ROM-G solutions are not as smooth as the true solutions,
particularly the velocity profile is very irregular, and the goal
of the closure model tuning is to smoothen both profiles, as much
as the closure model allows. For a clearer evaluation, we also
report the errors between the true solutions and the POD ROM-G
solutions, in Figure \ref{burgersstab_nome_test1_chap3}. Next, we
apply the ES-based closure model auto-tuning algorithm of Lemma
\ref{pdestab_lemma1_chap3}, and report the associated cost
function in Figure \ref{burgersstab_Q_test1lin_chap3}. We can
clearly see a decrease of the learning cost function over the
iterations, which means an improvement of the ROM prediction
performance. The corresponding profile of the linear closure model
amplitude over learning iterations is displayed in Figure
\ref{burgersstab_hatnue_test1lin_chap3}, which shows that it
reaches an optimal value around $\hat{\mu}_{e}\simeq 1.4$.  As
learning iterations proceed beyond the first $500$ displayed here,
the values remain stable around the same points. Thus, for clarity
of presentation, only a zoom around the first $500$ iterations is
displayed. Finally, the corresponding ROM solutions are shown in
Figure \ref{burgersstab_lromsol_test1lin_chap3}. For clearer
evaluation of the optimal closure model stabilizing effect on the
ROM, we also display in Figure \ref{burgersstab_le_test1lin_chap3}
the errors between the exact PDE solutions and the stabilized ROM
solutions.
\subsubsection{Test 2} 
\begin{figure}\center
  \begin{minipage}{0.4\linewidth}
   \center\subfigure[Learning-free POD ROM velocity profile]{
\includegraphics[width=1\linewidth]{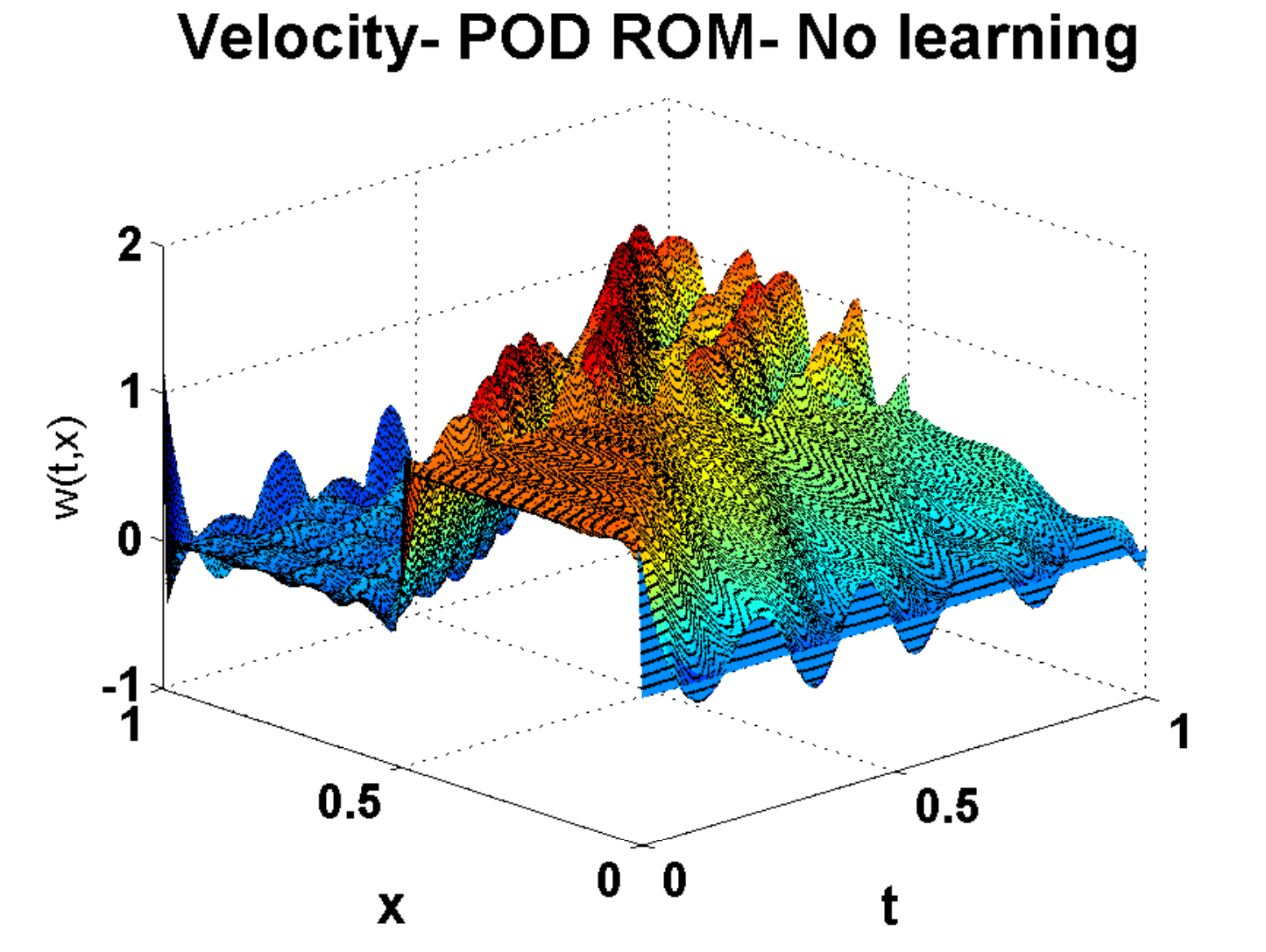}
\label{burgersstab_lfromv_test1_chap3}}
  \end{minipage}
  \hfill
  \begin{minipage}{0.4\linewidth}
   \center\subfigure[Learning-free POD ROM temperature profile]{
\includegraphics[width=1\linewidth]{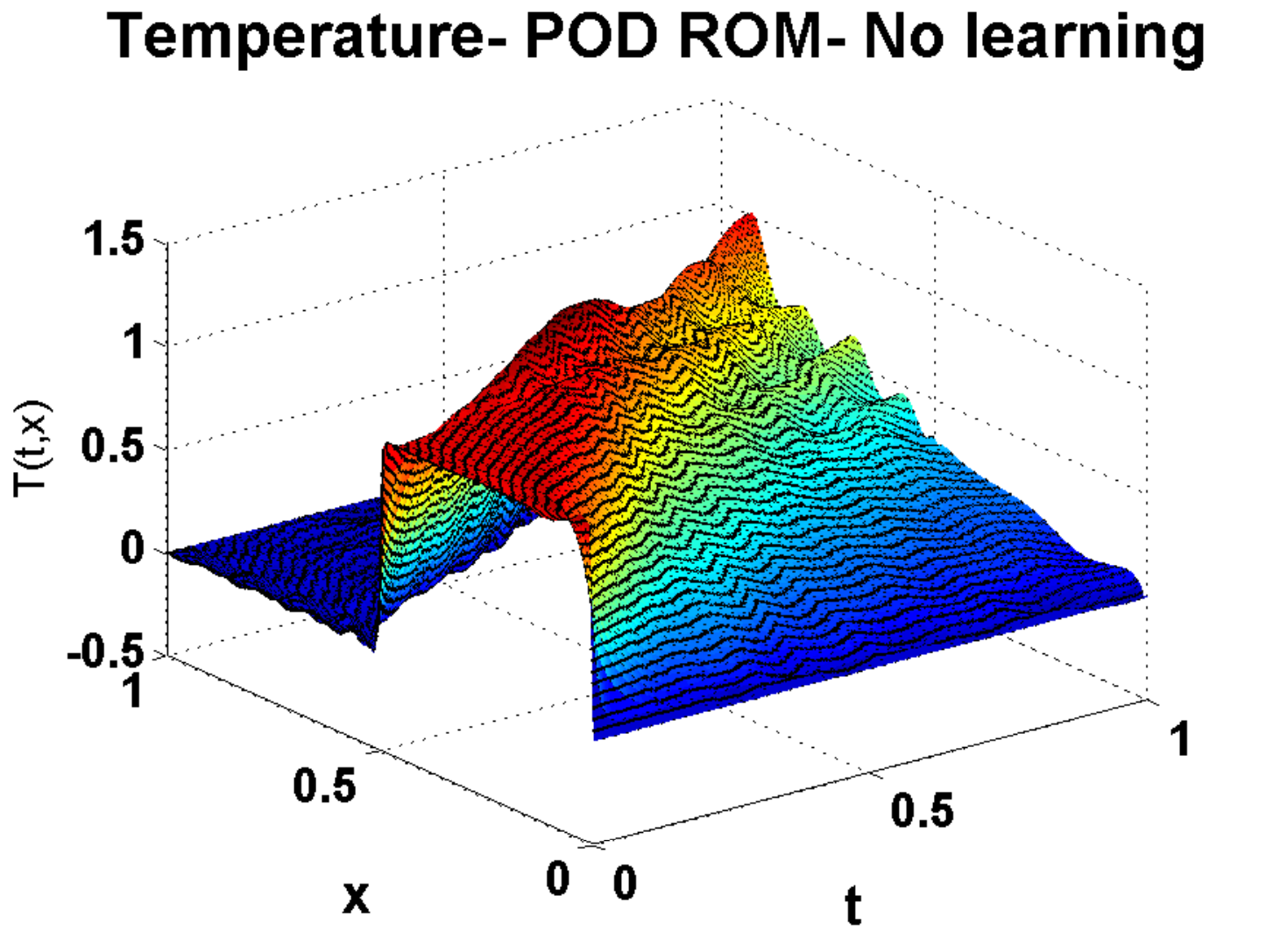}
\label{burgersstab_lfromt_test1_chap3}}
  \end{minipage}
  \caption{Learning-free POD ROM solutions of (\ref{Burgers2_chap3})- No learning}
  \label{burgersstab_lfromsol_test1_chap3}
\end{figure}

\begin{figure}\center
  \begin{minipage}{0.4\linewidth}
   \center\subfigure[Error between the true velocity and the learning-free POD ROM velocity profile]{
\includegraphics[width=1\linewidth]{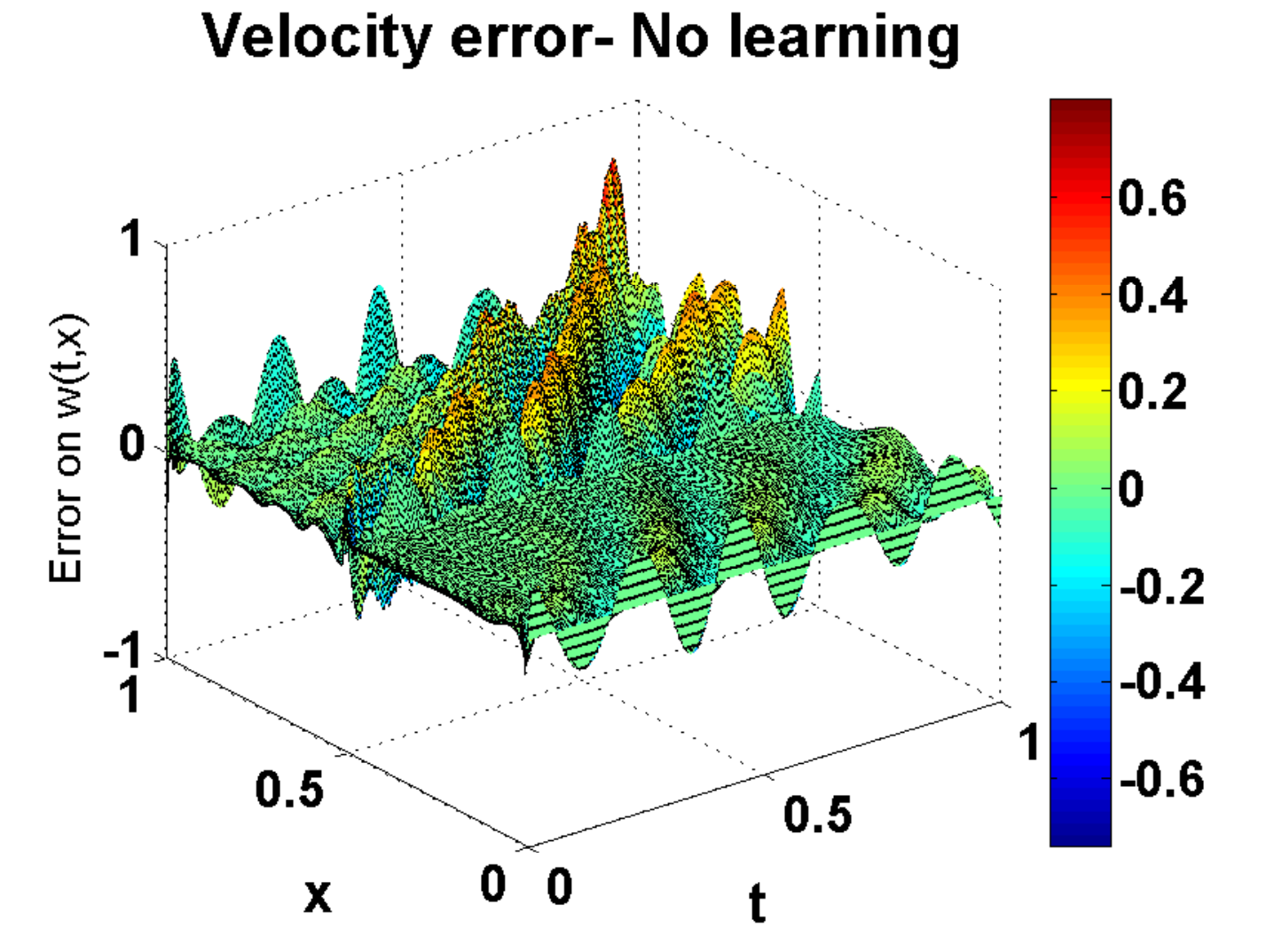}
\label{burgersstab_nomv_test1_chap3}}
  \end{minipage}
  \hfill
  \begin{minipage}{0.4\linewidth}
   \center\subfigure[Error between the true temperature and the learning-free POD ROM temperature profile]{
\includegraphics[width=1\linewidth]{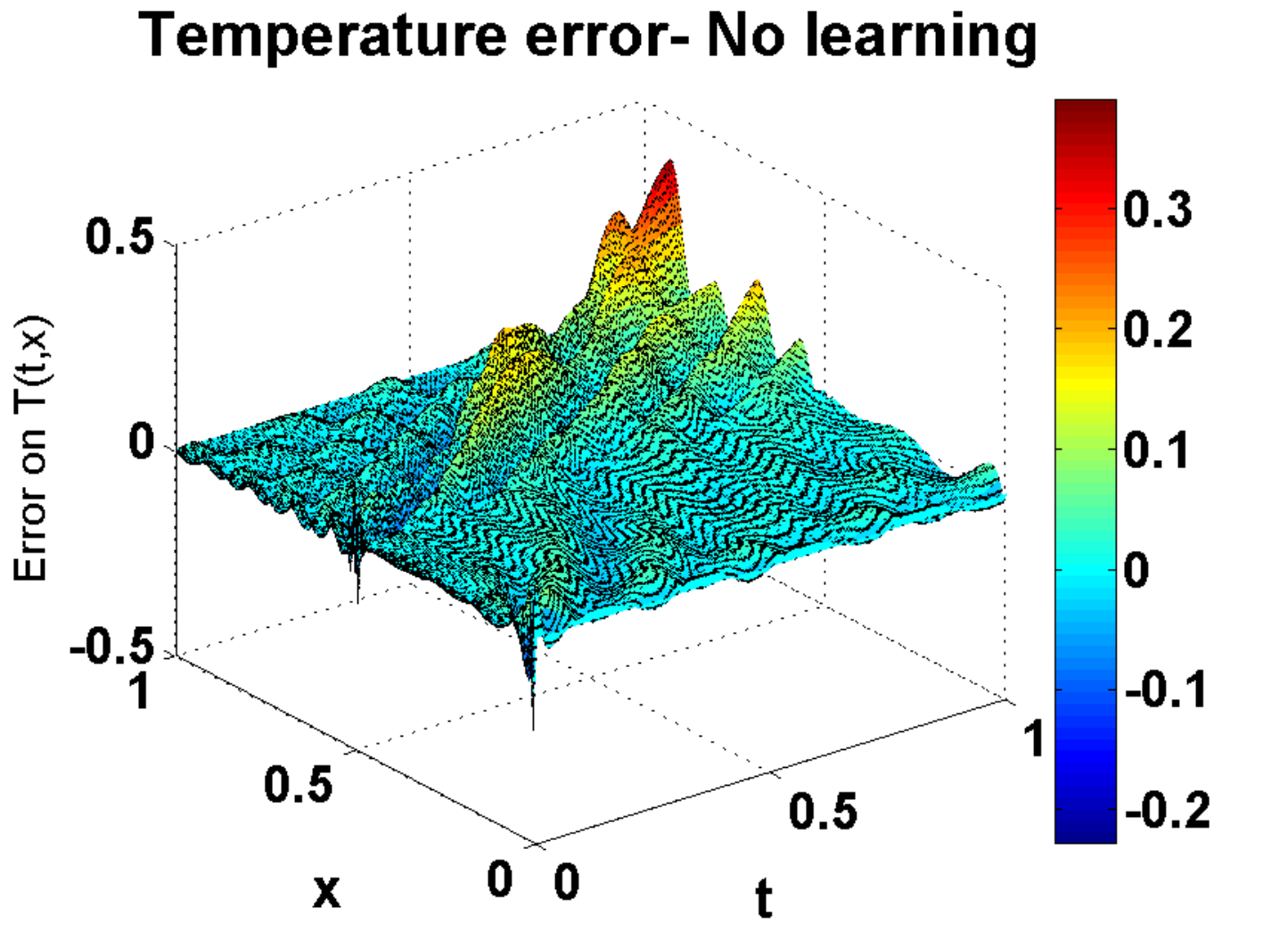}
\label{burgersstab_nomt_test1_chap3}}
  \end{minipage}
  \caption{Errors between the nominal POD ROM and the true solutions- No learning}
  \label{burgersstab_nome_test1_chap3}
\end{figure}

\begin{figure}\center
  \begin{minipage}{0.4\linewidth}
   \center\subfigure[Error between the true velocity and the learning-based POD ROM velocity profile]{
\includegraphics[width=1\linewidth]{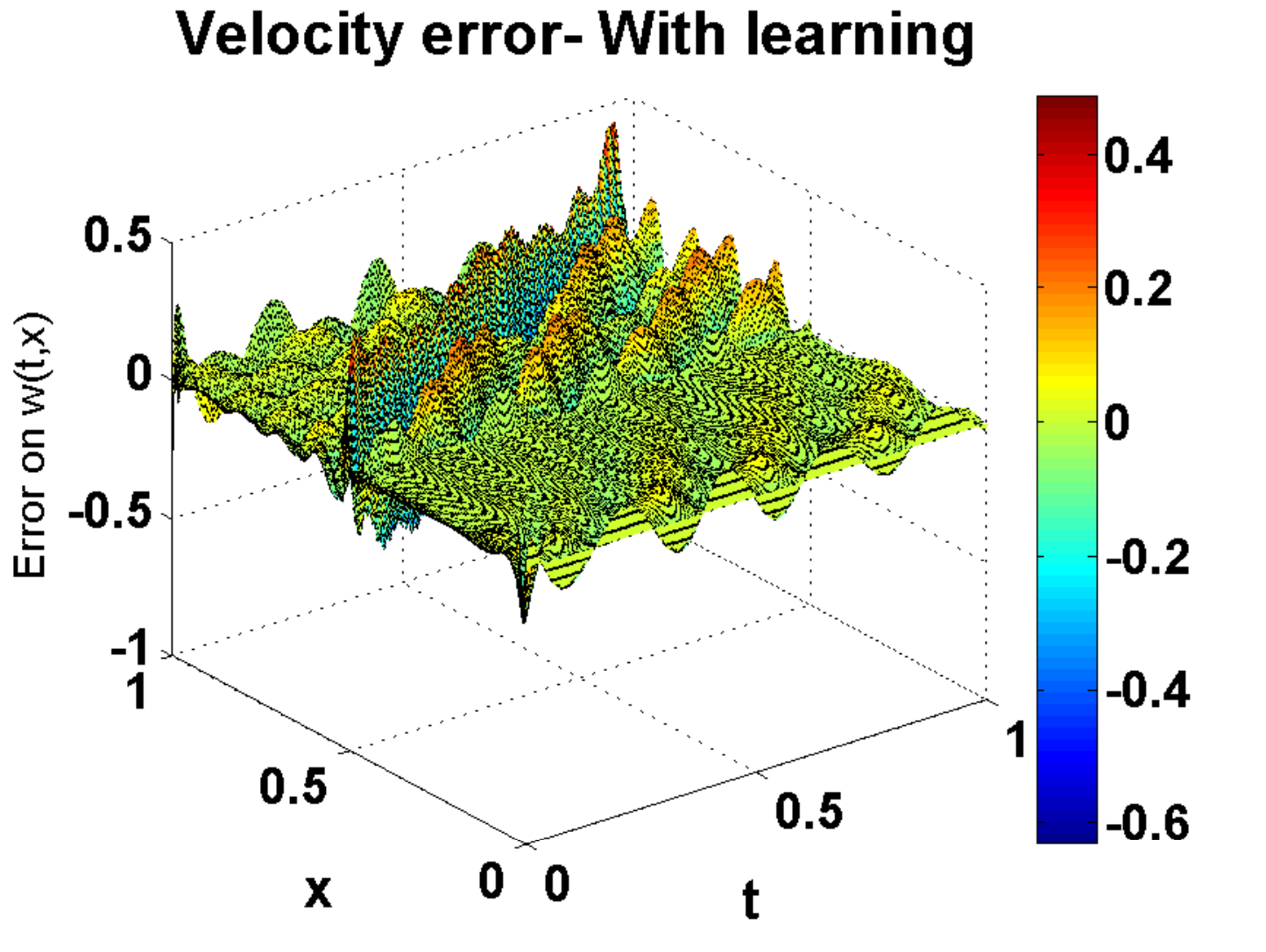}
\label{burgersstab_lev_test1lin_chap3}}
  \end{minipage}
  \hfill
  \begin{minipage}{0.4\linewidth}
\center \subfigure[Error between the true temperature and the
learning-based POD ROM temperature profile]{
\includegraphics[width=1\linewidth]{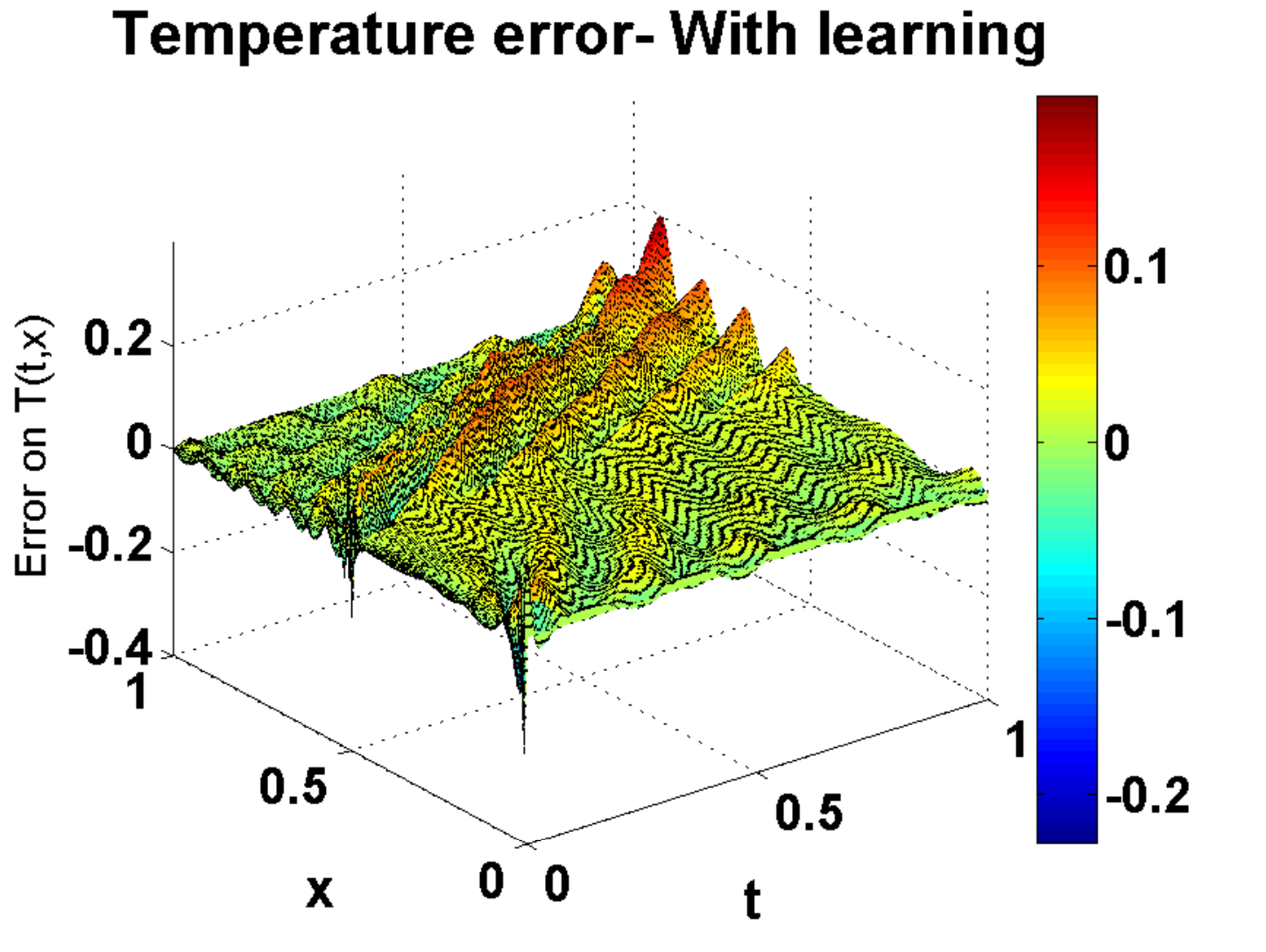}
\label{burgersstab_let_test1lin_chap3}}
  \end{minipage}
  \caption{Errors between the learning-based POD ROM and the true solutions- Stabilization with learning- Linear closure model}
  \label{burgersstab_le_test1lin_chap3}
\end{figure}

\begin{figure}\center
  \begin{minipage}{0.4\linewidth}
   \center\subfigure[Learning cost function vs. number of iterations]{
\includegraphics[width=1\linewidth]{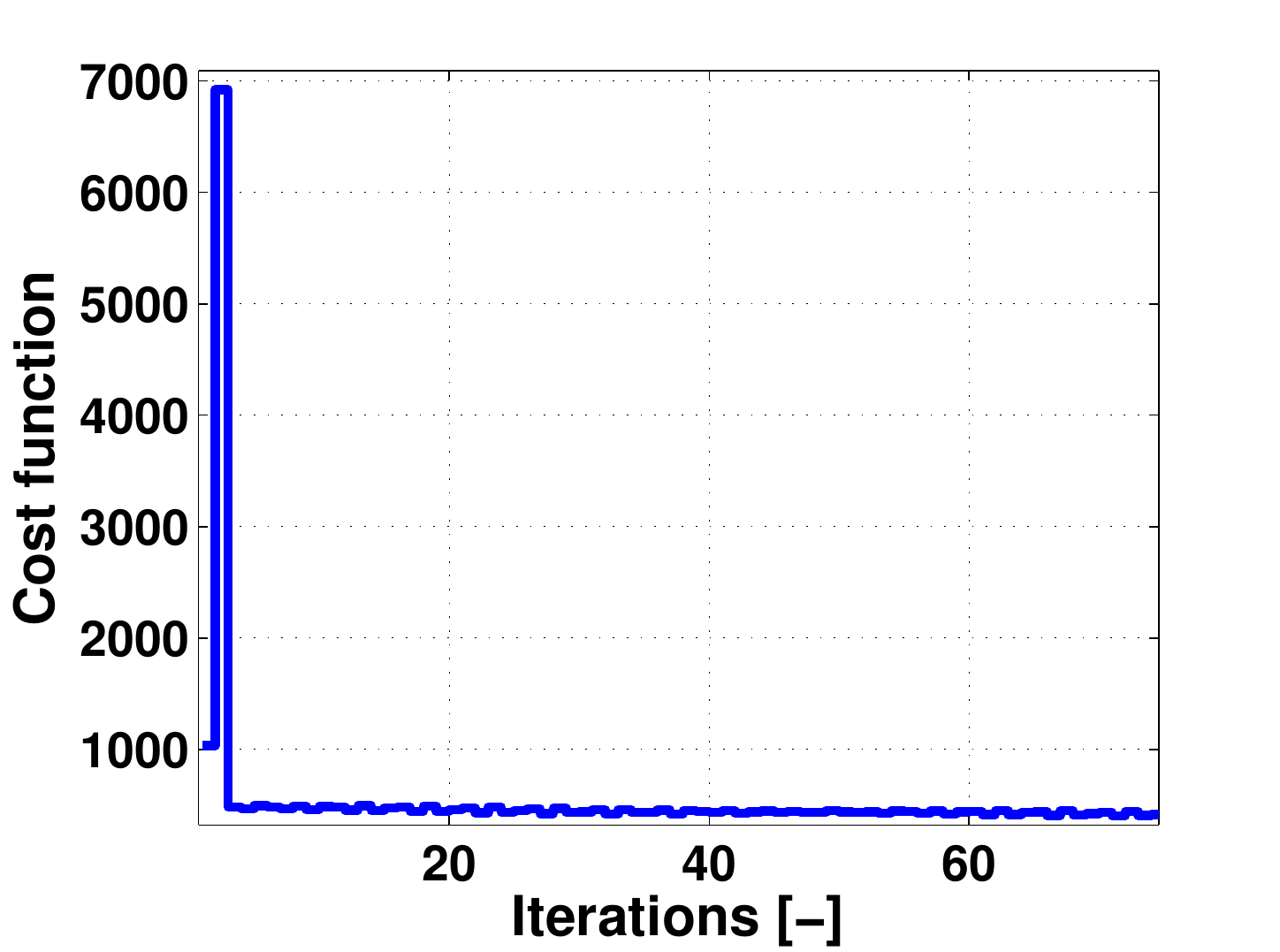}
\label{burgersstab_Q_test1_chap3}}
  \end{minipage}
  \hfill
  \begin{minipage}{0.4\linewidth}
   \center\subfigure[Learned parameter $\hat{\nu_{e}}$ vs. number of iterations]{
\includegraphics[width=1\linewidth]{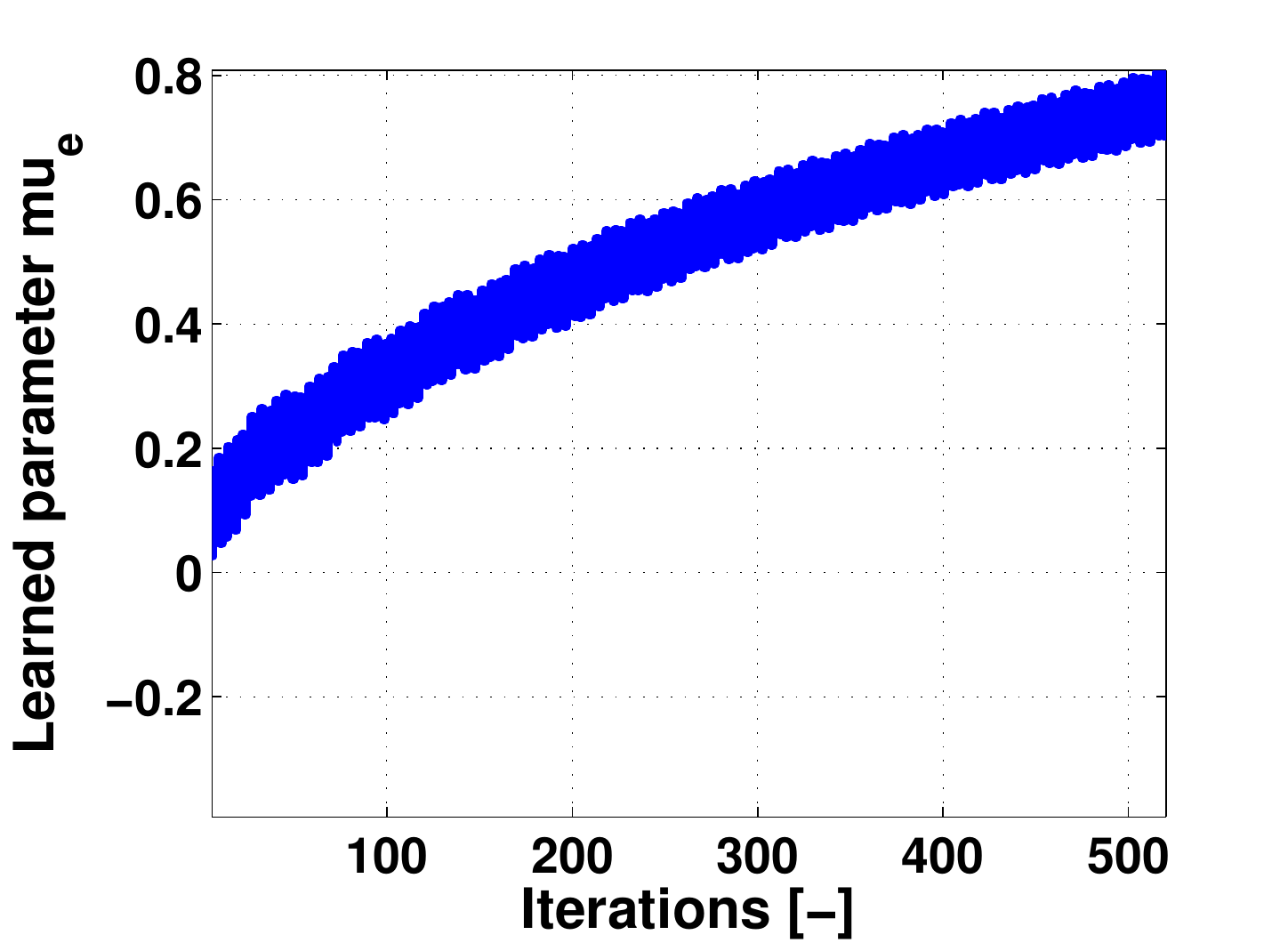}
\label{burgersstab_hatnue_test1_chap3}}
  \end{minipage}
  \center  \begin{minipage}{0.4\linewidth}
   \center\subfigure[Learned parameter $\hat{\nu_{nl}}$ vs. number of iterations]{
\includegraphics[width=1\linewidth]{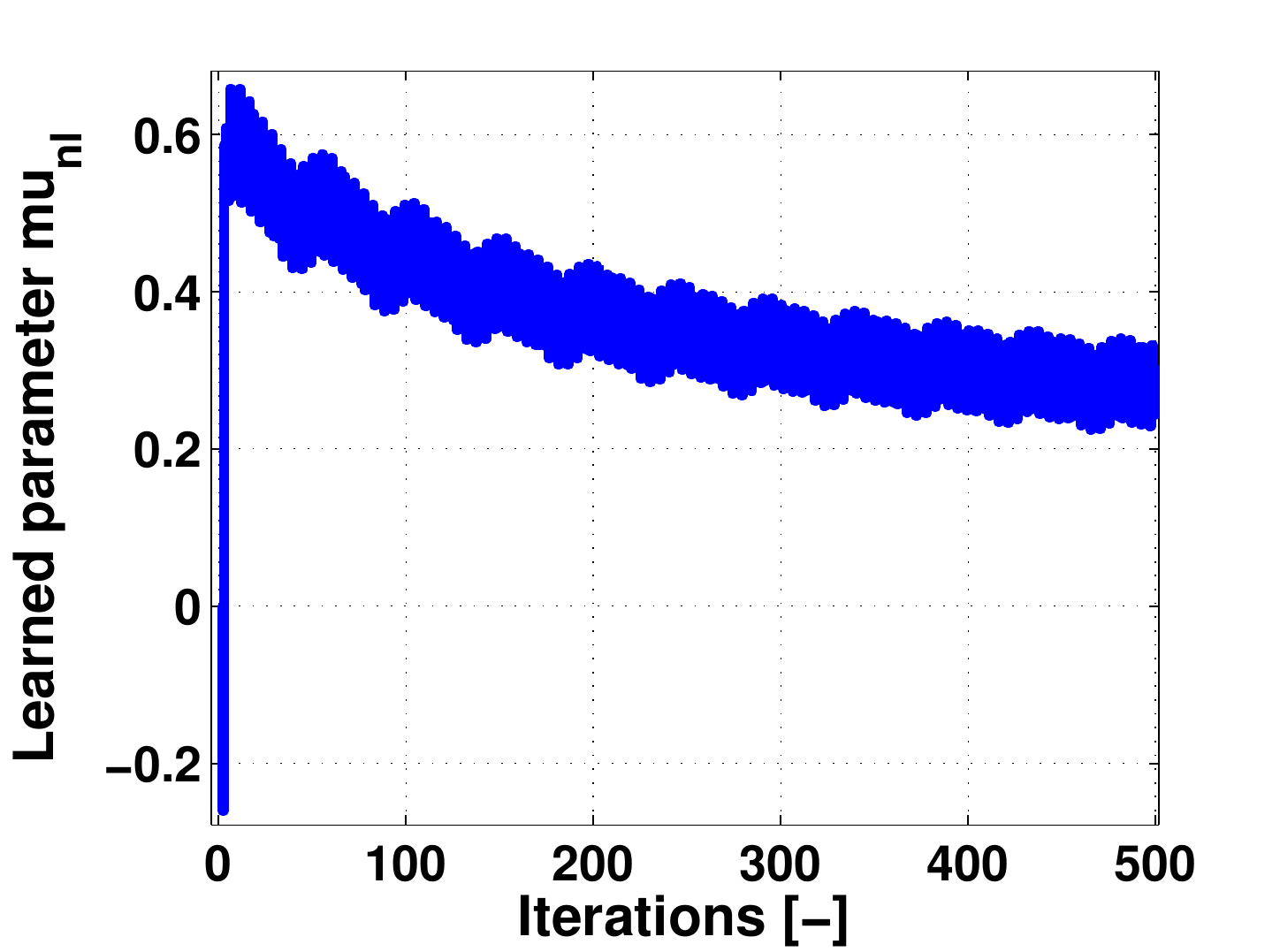}
\label{burgersstab_hatnunl_test1_chap3}}
  \end{minipage}
  \caption{Learned parameters and learning cost function- Stabilization with learning- Nonlinear closure model}
  \label{burgersstab_learningpara_test1_chap3}
\end{figure}

\begin{figure}\center
  \begin{minipage}{0.4\linewidth}
   \center\subfigure[Learning-based POD ROM velocity profile]{
\includegraphics[width=1\linewidth]{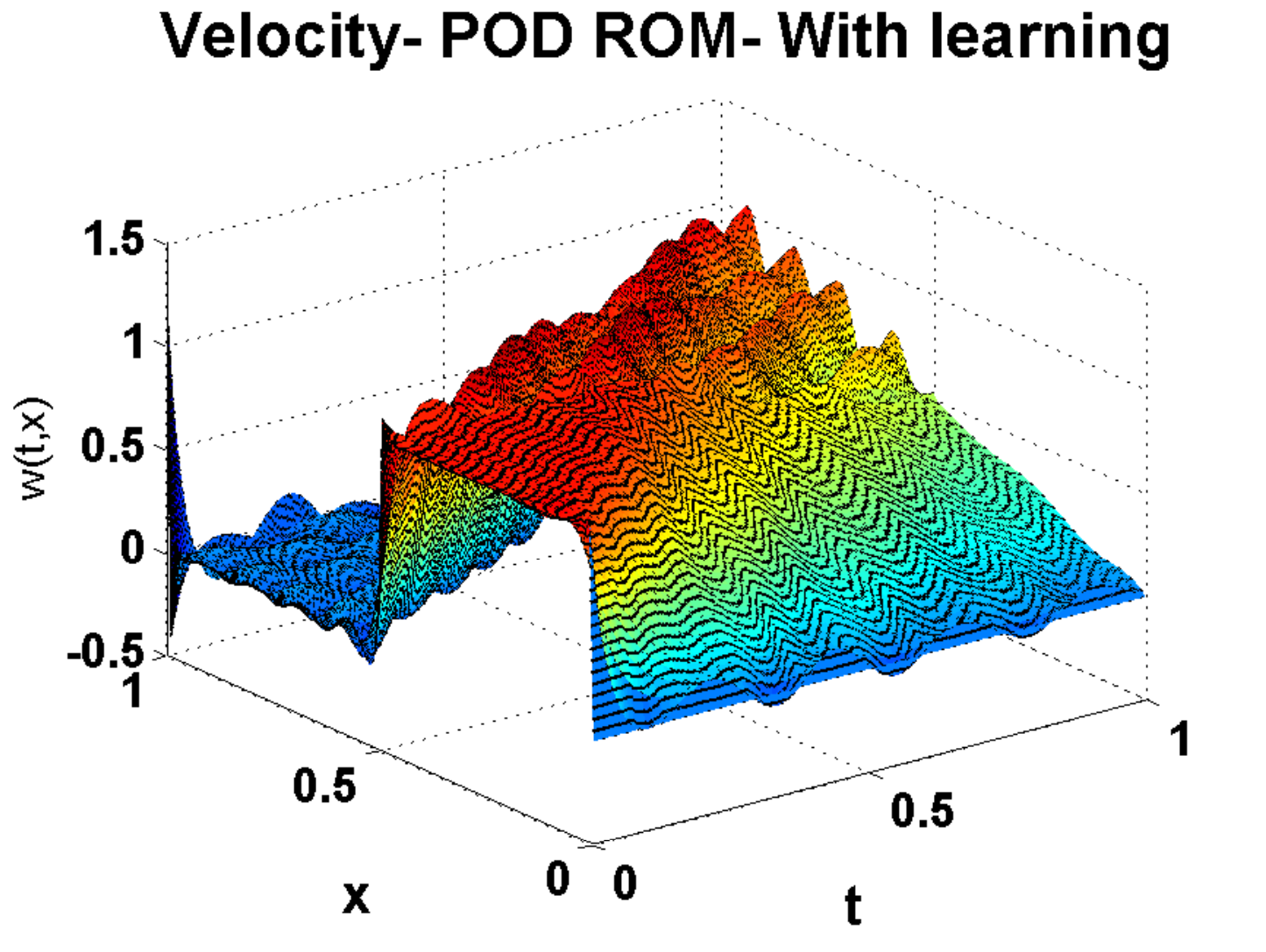}
\label{burgersstab_lromv_test1_chap3}}
  \end{minipage}
  \hfill
  \begin{minipage}{0.4\linewidth}
   \center\subfigure[Learning-based POD ROM temperature profile]{
\includegraphics[width=1\linewidth]{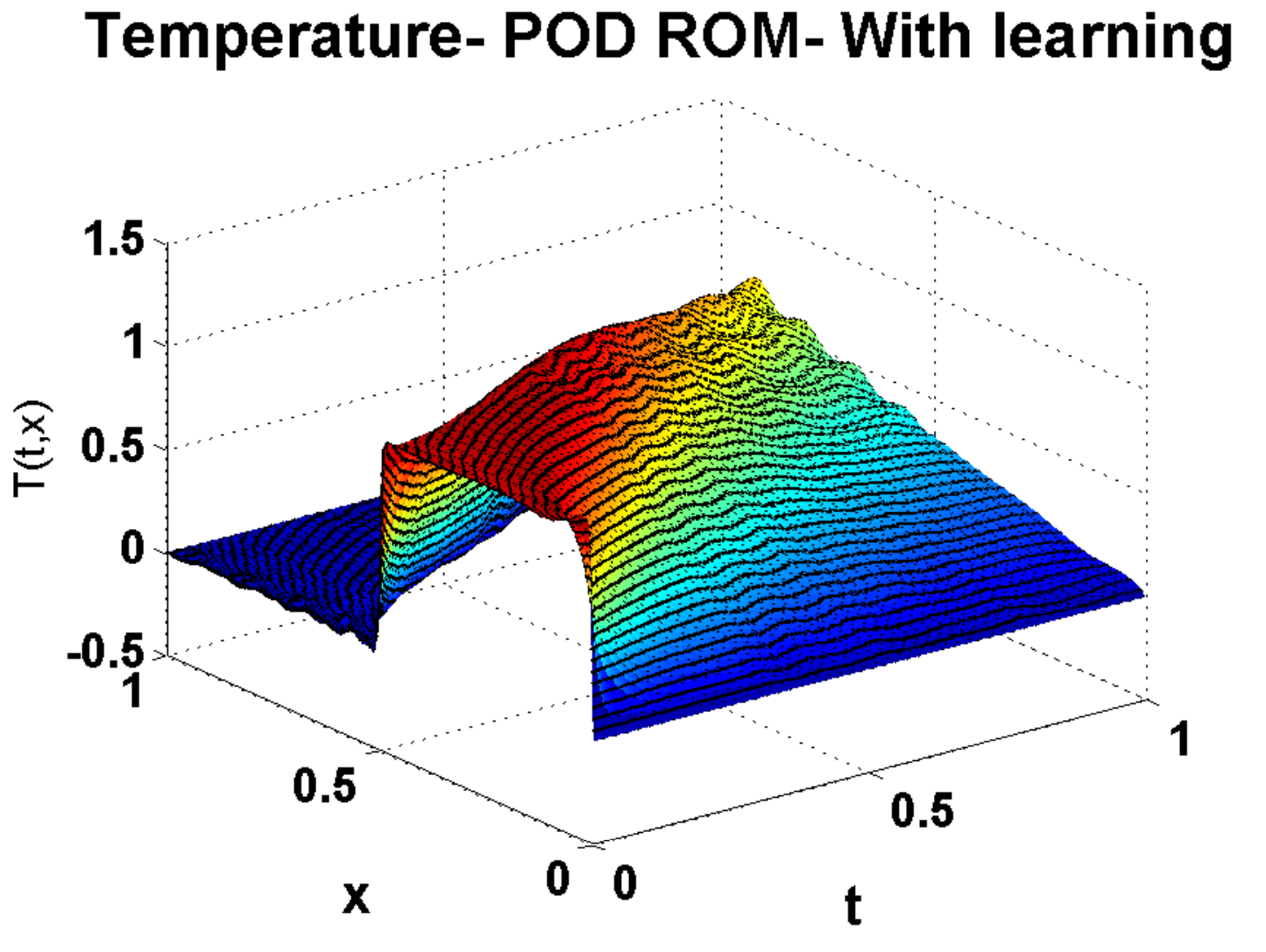}
\label{burgersstab_lromt_test1_chap3}}
  \end{minipage}
  \caption{Learning-based POD ROM solutions of (\ref{Burgers2_chap3})- Stabilization with learning- Nonlinear closure model}
  \label{burgersstab_lromsol_test1_chap3}
\end{figure}

\begin{figure}\center
  \begin{minipage}{0.4\linewidth}
   \center\subfigure[Error between the true velocity and the learning-based POD ROM velocity profile]{
\includegraphics[width=1\linewidth]{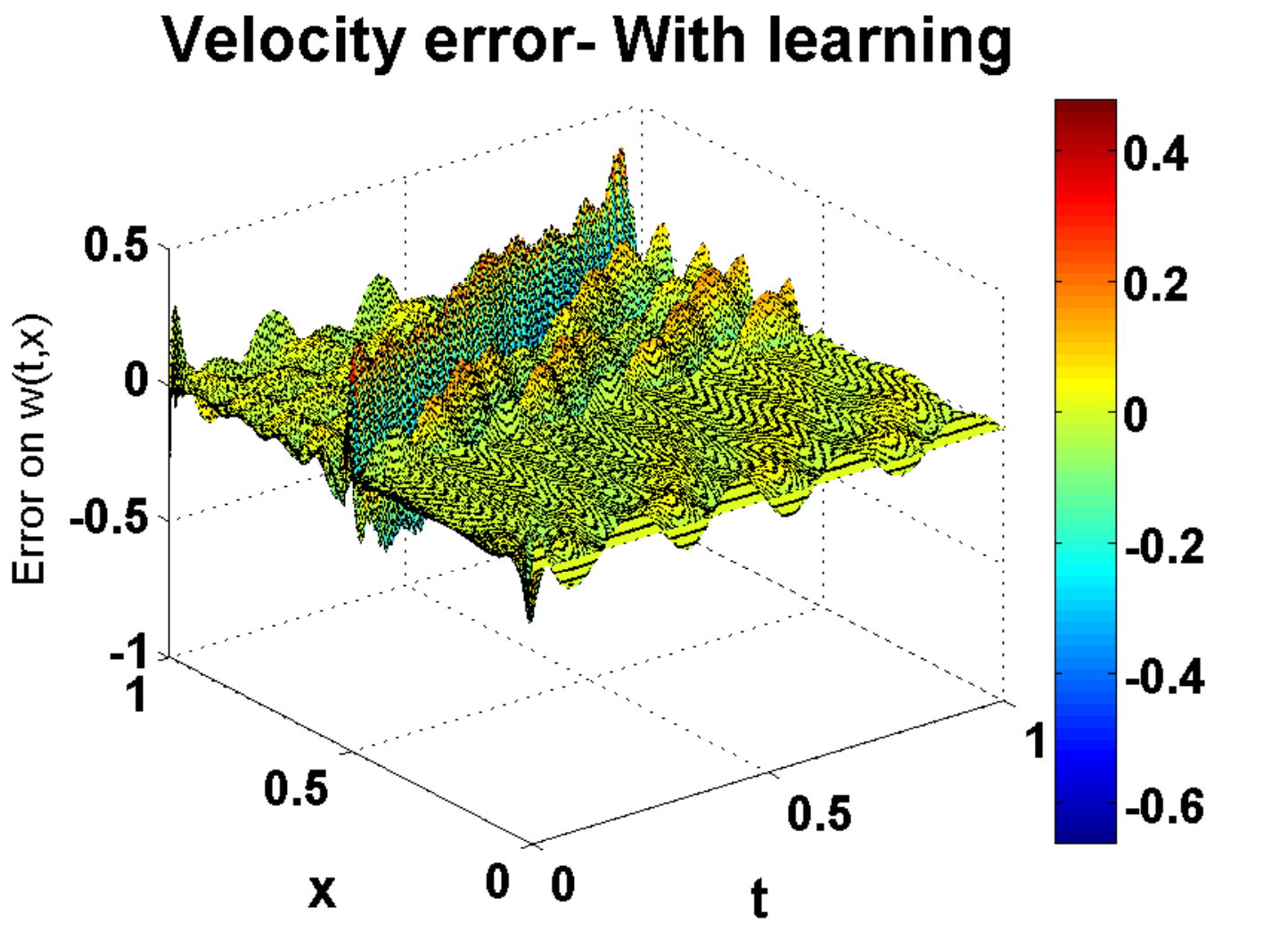}
\label{burgersstab_lev_test1_chap3}}
  \end{minipage}
  \hfill
  \begin{minipage}{0.4\linewidth}
\center \subfigure[Error between the true temperature and the
learning-based POD ROM temperature profile]{
\includegraphics[width=1\linewidth]{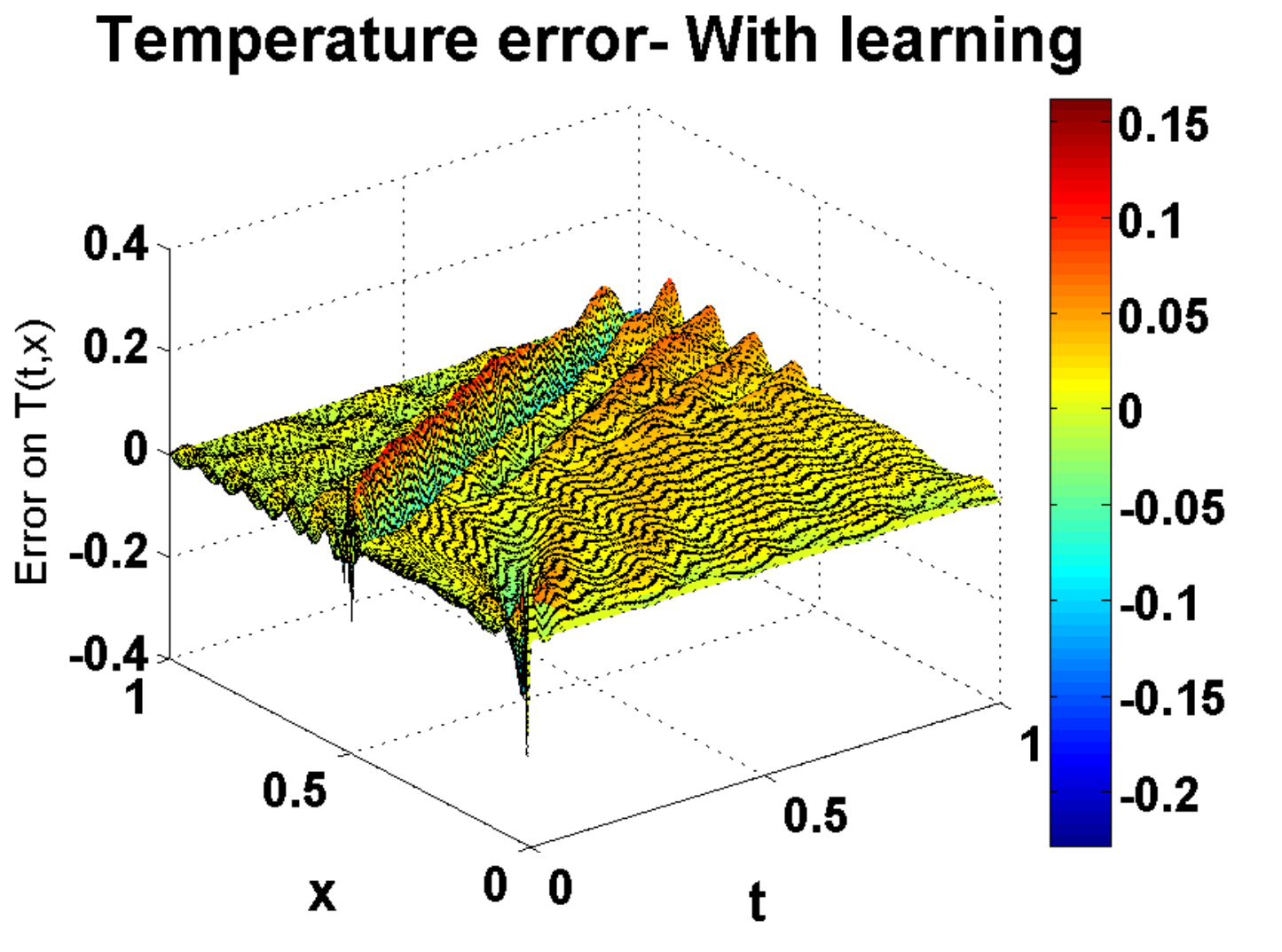}
\label{burgersstab_let_test1_chap3}}
  \end{minipage}
  \caption{Errors between the learning-based POD ROM and the true solutions- Stabilization with learning- Nonlinear closure model}
  \label{burgersstab_le_test1_chap3}
\end{figure}

We report here the case related to Lemma
\ref{pdestab_lemma2_chap3}, where we test the auto-tuning results
of ES on the combination of a linear constant viscosity closure
model and a nonlinear closure model. We apply Lemma
\ref{pdestab_lemma2_chap3}, with the Heisenberg linear closure
model given by (\ref{PODROMH_CHAP3}). The two closure model
amplitudes $\mu_{e}$ and $\mu_{nl}$ are tuned using the discrete
version of the ES algorithm (\ref{pdestab_mes_2_chap3}), given by
\begin{equation}
\begin{array}{l}
y_{1}(k+1)=y_{1}(k)+a_{1}\;t_{f}\;\sin(\omega_{1} t_{f}k+\frac{\pi}{2})Q(\hat\mu_{e},{\hat\mu}_{nl}),\\
\hat{\mu}_{e}(k+1)=y_{1}(k+1)+a_{1}\;\sin(\omega_{1} t_{f}k-\frac{\pi}{2}),\\
y_{2}(k+1)=y_{2}(k)+a_{2}\;t_{f}\;\sin(\omega_{2} y_{2}(k+1)+\frac{\pi}{2})Q(\hat\mu_{e},{\hat\mu}_{nl}),\\
\hat{\mu}_{nl}(k+1)=y_{2}(k+1)+a_{2}\;\sin(\omega_{2}
t_{f}k-\frac{\pi}{2}), \label{pdestab_mes_2discrt_chap3}
\end{array}
\end{equation}
where $y_{1}(0)=y_{2}(0)=0$, and $k=0,1,2,...$ is the number of
the learning iterations. We use the parameters' values:
$a_{1}=6\times 10^{-6}\;[-],\;\omega_{1}=10\;
[\frac{rad}{sec}],\;a_{2}=6\times
10^{-6}\;[-],\;\omega_{2}=15\;[\frac{rad}{sec}]$, and a similar
cost function as in Test 1. We show the profile of the learning
cost function over the learning iterations in Figure
\ref{burgersstab_Q_test1_chap3}. We can see a quick decrease of
the cost function within the first $20$ iterations. This means
that the ES manages to improve the overall solutions of the POD
ROM very fast. The associated profiles for the two closure models'
amplitudes learned values $\hat{\mu}_{e}$ and $\hat{\mu}_{nl}$ are
reported in figures \ref{burgersstab_hatnue_test1_chap3}, and
\ref{burgersstab_hatnunl_test1_chap3}. We can see that even though
the cost function value drops quickly, the ES algorithm continues
to fine-tune the values of the parameters $\hat{\mu}_{e}$,
$\hat{\mu}_{nl}$ over the iterations, and they reach eventually
reach an optimal values of $\hat{\mu}_{e}\simeq 0.3$, and
$\hat{\mu}_{nl}\simeq 0.76$. We also show the effect of the
learning on the POD ROM solutions in Figure
\ref{burgersstab_lromsol_test1_chap3}, and figure
\ref{burgersstab_le_test1_chap3}, which by comparison with figures
\ref{burgersstab_lfromsol_test1_chap3}, and
\ref{burgersstab_nome_test1_chap3}, show a clear improvement of
the POD ROM solutions with the ES tuning of the closure models'
amplitudes. We also notice an improvement of the ROM solutions
compared to the linear closure model stabilization test results of
figure  \ref{burgersstab_le_test1lin_chap3}. Specifically, we see
that the temperature error in the case of the closure model of
Lemma \ref{pdestab_lemma2_chap3}, is smaller than the one obtained
with the linear closure model of Lemma \ref{pdestab_lemma1_chap3}.
We do not currently have a formal proof. However, we believe that
the improvement is due to the fact that in the closure model of
Lemma \ref{pdestab_lemma2_chap3}, we are using both the
stabilizing effect of the linear viscosity closure model term and
the stabilizing effect of the nonlinear closure model term.



\newpage
\section{Conclusion and discussion}\label{concl}
In this work, we explore the problem of stabilization of reduced order
models for partial differential equations, focusing on the closure
model-based ROM stabilization approach. It is well known that tuning
the closure models' gains is an important part in obtaining good
stabilizing performances.  Thus, we propose a learning ES-based
auto-tuning method to optimally tune the gains of linear and nonlinear
closure models, and achieve an optimal stabilization of the ROM. We
validate our using the coupled Burgers' equation as an example,
demonstrating significant gains in error performance. The results are
encouraging. We defer to future publications verifying our approach on
more challenging higher dimensional cases. Our results also raise the
prospect of developing new nonlinear closure models, together with
their auto-tuning algorithms using extremum seeking, as well as other
machine learning techniques.

\bibliographystyle{IEEEtran}
\footnotesize{\bibliography{bibliov3}}

\end{document}